\documentclass{emulateapj}

\slugcomment{Draft; Version: 11 September 2006}

\shorttitle{CMa stellar over-density}
\shortauthors{Butler et al.}

\begin{document}

\title{A Canis Major over-density imaging survey. I. Stellar content and star-count maps:\\
 A distinctly elongated body of  main sequence stars}

\author{D. J. Butler$^1$, D. Mart\'inez-Delgado$^{1,2,3}$, H-W. Rix$^1$,
  J. Pe\~narrubia$^{1,4}$, J. T. A. de Jong$^1$}
\affil{$^1$Max-Planck-Institut f\"ur Astronomie, K\"onigstuhl 17, D-69117
  Heidelberg, Germany}
\affil{$^2$Instituto de Astrof\'isica de Andalucia (CSIC), Granada,
Spain} 
\affil{$^3$Instituto de Astrof\'isica de Canarias, C/ V\'ia L\'actea,
 E38200 - La Laguna, Spain}
\affil{$^4$Department of Physics and Astronomy, 
 University of Victoria, 3800 Finnerty Rd., Victoria, BC, V8P 5C2, Canada}
\email{butler@mpia.de}

\begin{abstract}
 We present  first results from a large-area 
 ($\sim$80$^\circ$$\times$20$^\circ$), sparsely sampled
  two-filter (B,R) imaging survey  
 towards the Canis Major stellar over-density, 
  claimed to be a disrupting Milky Way satellite galaxy.
  Utilizing stellar colour-magnitude
  diagrams reaching to B$\sim$22\,mag, 
 we provide a first delineation of its 
 surface density distribution using main sequence stars.
  It is located below the  Galactic mid-plane,  and  
 can  be discerned to at least b=-15$^\circ$. 
  Its projected shape is highly elongated, nearly parallel to the
  Galactic plane, with an axis ratio of $\ga$ 5:1, substantially more
  so than what Martin et  al. originally found.  
 We  also provide a first map of 
 a prominent over-density of blue, presumably younger main sequence
   stars, which  appears to have a maximum  
  near [l,b $\sim$ 240$^\circ$, -7$^\circ$ ]
   and extends in latitude to b$\sim$ -10$^\circ$.
  The young population  is markedly more localized. 
  We estimate an {\it upper} limit  on the  line-of-sight depth,
 $\sigma_{\rm los}$,
  of the old population based on the main sequence width, by  comparing
  with a simple stellar population (Pal 12), obtaining 
  $\sigma_{\rm los}$  $<$ 1.8 $\pm$ 0.3\,kpc,
  at an adopted  D$_\odot$ = 7.5 $\pm$ 1\,kpc.  
 For the young 
  stellar population, we find  $\sigma_{\rm los}$ 
  $<$ 1.5\,kpc. The overall picture
 presented is one of a young stellar population that is less extended,
 both in terms its line-of-sight depth and angular size, than the
  older population. 
 In the literature  three  explanations for the 
  CMa over-density have been invoked,
  namely (a) a partially disrupting dwarf galaxy on
  a low-latitude orbit,  (b) a projection of a 
  warped outer Galactic disk, and (c) a projection of 
  an out-of-plane spiral arm. 
  While the data provide no firm arguments against the less well-defined
  third scenario,  they have clear implications for 
  each of the  others: (a) We infer from the strong elongation of the
  over-density in longitude, and simulations in the literature, that
  the CMa over-density  is unlikely to be a gravitationally bound
  system at the present  epoch, but may well be just a recently disrupted
  satellite remnant. A complexity for the satellite origin
  may arise from the `flattening'
  of the young MS population, which is possibly more pronounced than
  the older one.  (b) Based on modeling, we find that the 
  line-of-sight depth of the MS over-density in old stars is clearly
  inconsistent with published locally axi-symmetric descriptions 
  of the warped Galactic disk, such as those considered in
  Momany et al. (2006). Without detailed modeling, the data-set
  itself does not provide sufficient leverage to distinguish between
  an interpretation as  sub-structure in the warped outer Galactic  disk
  or a disrupted satellite. 
 \end{abstract}

\section{Introduction}\label{intro}
The current paradigm in which  large disk galaxies like our Milky Way form
 (e.g. White \& Rees 1978; White \&  Frenk 1991),
 is based on the successive coalescence and accretion of
 smaller systems of dark matter,  stars,  and
 diffuse interstellar  gas into larger assemblies.
 Some of these smaller systems may be satellite dwarf galaxies that can 
  be disrupted 
   by tidal shocks and evaporation in a Galactic potential, and
 can spawn tidal tails.
 Numerical simulations exist that model
  distinct tidal stellar streams in and around
large galaxies, indicating that at r$\ga 10$kpc such streams 
 should remain detectable as coherent
stellar over-densities for billions of years (Johnston et al. 1999;
 Ibata \& Lewis 1998;  Mart\'inez-Delgado et al. 2004;
  Pe\~narrubia et al. 2006).

Recently, much work 
 has focused on a possible low-latitude 
 stellar stream, the so-called Monoceros stream, 
 encircling the Milky Way (MW) at a galacto-centric distance of around 
 20\,kpc, which was
 found through colour-magnitude diagrams (CMDs) from SDSS\footnote{\it Sloan
  Digital Sky Survey} (Newberg et al. 2002; Yanny et al. 2003); also  
 see  Ibata et al. (2003),   Conn et al. (2005) \& Martin et al. (2006).
 For stellar over-densities at low galactic latitudes it is 
not obvious  a priori whether such a stream is of external origin, i.e. 
the tidal debris of  a now (partially) disrupted  
satellite or, alternatively, a distorted part of the pre-existing
outer stellar disk.

Arguments in favour of the dwarf satellite hypothesis have come from
 dynamical modeling (Pe\~narrubia et al. 2005), showing that it is
 possible to find a plausible model  of the Monoceros stream 
 that explains all detected parts  
 of this low-latitude stream  as the wrapped tidal debris
 tails of a disrupting (model) dwarf galaxy. In this dynamical 
 model the position of
 the CMa over-density is  not an initial constraint, and yet  
 the main body of the disrupting model satellite happens to 
 located in the direction of CMa at the present epoch. 
  All kinematics, including
 the subsequently determined proper motion (Dinescu et al. 2005) 
 of main sequence (MS) stars 
 in a small area (0.25\,deg$^2$) towards the CMa stellar over-density 
 are consistent with the  dynamical model. Additional empirical support
  appears to come from the apparent 
 bifurcation of suspected Monoceros stream stars at right
 ascension $\sim$ 120$^\circ$ 
   in Fig.1 of Belokurov et al. (2006), as could be expected
 in general based on the above-mentioned 
 Pe\~narrubia et al. model.  Other evidence  comes
 from the discovery of an over-density of 
 possible  Monoceros stream RR Lyrae stars (Vivas et al. 2006),
  whose distances closely match the model expectation in 
  Pe\~narrubia et al. (2005).

The position and survival of
the  progenitor of this stellar stream is still under debate. The best
candidate is a seemingly well-defined stellar over-density of stars 
 discovered in the direction of Canis Major (CMa) using 2MASS red
 giants (Martin et al. 2004a). This discovery has spawned a lively  
 debate in the literature on whether this apparent 
 over-density of red giants could be 
 part of the distorted outer stellar disk (Momany et
 al. 2004; 2006; see also Rocha-Pinto et al. 2006) 
  or an accreted and possibly disrupting satellite  [Martin et
 al. 2004b; Martinez-Delgado et al 2005 (MD05); Bellazzini et al. 2006 (B06)]. 
 However, a firm large-area kinematical, spatial and chemical link between
 the three known stellar  components of the overall debate 
 (i.e. the young and old stellar over-densities, and the Monoceros
 stream)  is lacking.

From deep, visible-band photometry there 
 are signs of at least two different star formation episodes
 in the direction of CMa
  [Bellazzini et al. (2005), B05; MD05; Carraro et al. (2005)]. For the sake of clarity, we
  illustrate this in an annotated colour-magnitude diagram,
  Fig.~\ref{Fig_Select_CMD_single}, which corresponds to a field 
 near the presumed centre of the CMa
 over-density, and where the data are taken from this survey (see
 Sec.~\ref{obs_datared}).      Whether the most recent burst  
 of star formation activity occurred 1-2\,Gyr ago, as touted in B05,
  has been debated, in favour of 
  a much younger stellar population
  of $\la$ 100\,Myr-old stars (Carraro et al. 2005).
  The CMa over-density  itself, which is taken to lie near
  D$_\odot$=7.5\,kpc
  (B05,MD05), comprises a predominantly 
 older population of stars (4-10\,Gyr; B05)
 and Moitinho et al. (2006) suggest that it may be 
  consequence of viewing a  local
 arm-like MW stellar sub-structure in projection.
  For a different and independent study of the stellar populations,
  we will make use of detailed CMD-fitting 
  in a forth-coming paper (de Jong et al., in prep.).

 To determine the full angular extent of the young and old main
 sequence stars towards CMa 
 and to test whether they stem from a satellite galaxy,   
 we conducted a large-area ($\sim$80$^\circ$$\times$20$^\circ$),
 sparsely sampled visible-band imaging survey of the CMa region,
 drawing on the wide field imager (WFI) at the MPG/ESO 2.2\,m
 telescope on La Silla. 
 The survey sub-area considered in this
 paper is 230$^\circ$ $\la$ l  $\la$ 260$^\circ$, -20$^\circ$ $\la$ b
 $\la$ 15$^\circ$.

\subsection{Aim and design of the WFI survey}
The principal goal of our
 imaging survey is to provide a database that can 
  address ultimately whether the CMa
stellar over-density results from a dynamical distortion of the outer
 Galactic disk or whether it is  
 the remains of a formerly large 
 dwarf galaxy.  As a step towards this goal, we aim to 
  map its  full angular extent through its star-count
  profile.

 Recent reports (B06) based on 2MASS red clump  stars 
 suggest that the discernable part of the CMa over-density
 may extend across several 100deg$^2$ on the sky.
  Our survey goal is to cover 
 this over-density through sparse sampling, but with relatively deep
 visible-band imaging. Specifically, we want to obtain
  CMDs that reach  well below the turn-off magnitude of 
 an ancient stellar population; our target limiting magnitude is B, R
 = 23\,mag (S/N = 10). We chose a wide 
 colour baseline (B-R) to have 
  sensitivity to age- and/ or [Fe/H]-dependent
 CMD features
  (young MS, old MS turn-off and the red clump). 
 The key difference (and merit)
  of the present CMa survey over previous large-area
 surveys, which have
  been based exclusively on 2MASS red-giant stars and MW models, is
 that we base our analysis on 
 colour-magnitude diagrams that exhibit 
 a well-defined stellar main sequence at a defined distance range. This
  provides an independent and  high-contrast way of tracing
 the full  angular extent of the  CMa over-density. The chief
 disadvantage of visible-band data is that Galactic extinction in many
 parts of this area is high, A$_{\rm V}$$\ga$0.5\,mag.

\subsection{Aim of this paper}
In this paper we present the 
 data from the first observing phase of our imaging survey,
  covering the region around [l,b = 240$^\circ$, -8$^\circ$]
 that had originally been identified as
 the over-density in Martin et al. (2004a).
   As it turns-out, these data provide only 
 partial spatial coverage of the over-density.   
 Therefore, we  refrain from a wide-spread quantitative comparison
  with Galaxy star-count models, except  (i) to
 use a Besancon model CMD as a visual guide for the reader
 to help interpret the general morphology in our control CMDs, and
 (ii) to aid the discussion in Sec.~\ref{interp}.

 We report our observations, and present the data and their reduction
 in Sec.~\ref{obs_datared}.
 The issue of photometry completeness and uncertainties 
 is assessed in Sec.~\ref{complt_tst}. The crucial issue of
 reddening by dust extinction
 and its correction is presented in Sec.~\ref{extinc}. In
 Sec.~\ref{controlfields}, we explain the choice of control fields,
 which  is followed by a description of the
 stellar content and morphology of the CMDs in
  Sec.~\ref{stel_content}. 
 In   Sec.~\ref{ms_counts} we provide a    
 star-count analysis of young and old CMa main sequence stars.
We estimate an upper limit on the
  line-of-sight size of the CMa stellar overdensity in Sec.~\ref{los}.
 We briefly discuss our results in  Sec.~\ref{discussion} 
 and  summarize the key results in Sec.~\ref{conclusions}. 

\section{Observations, data and data reduction}\label{obs_datared}
All observations were carried out in service 
mode with the wide-field imager (WFI) on 
 the 2.2-m ESO/MPG telescope at the La
 Silla observatory (Chile) from December 9 to 20,
 2004.  The WFI's field-of-view covers 0.25\,deg$^2$, sampled at
 0.238$^{\prime\prime}$ pixel$^{-1}$.
 Single exposure images 
 were taken in the B- and R-bands, at 100\,s per pointing and filter.
  A total of 59 pointings  (Table~\ref{survey_params})
 provides a sparse map of the stellar over-density
 region originally identified in Martin et al. (2004a).  
 B- and R-band pointings typically differ by less than 2.5\,arcsec,
 which leads to  negligible field-to-field differences in the (effective)
 imaging area.
 Overscan, bias, flat-field corrections, and  an astrometric correction
  were performed using a pre-reduction pipeline (Schirmer et
 al. 2003). We obtained stellar photometry using 
 DAOPHOT tools, available in IRAF.\footnote{Image
   Reduction and  Analysis Facility.} For each WFI pointing, we employed
  point spread
 function (PSF)-fitting using a spatially variable (order 2) moffat
 function of exponent 2.5, derived from 100 stars.
 The DAOPHOT/ALLSTAR PSF-fitting task is run on each frame separately.
As we have single exposures
  per pointing,    cosmic rays are effectively
 filtered out during the PSF-fitting and  the matching of the B- and R-band 
 photometry lists. 
  Objects in each list are then assigned Galactic coordinates
  and are  matched using a separation tolerance of 1.7\,arcsec.
 This large tolerance ensures that the faintest stars, 
  which have the poorest positional errors, are matched; and is
  tolerable owing to the  sparseness of the fields surveyed.

For the photometric zero-points, we firstly
  tie the photometry from each observing night to  
 a common  internal (i.e. instrumental) zero point using the 
 zero-point of each PSF (computed from its flux and magnitude).
 This is followed by 
 the   transformation of the instrumental magnitudes
 into the 
 standard Johnson-Cousins photometric system based
 on observations  of either 
 standard stars from Landolt (1992) [for Dec. 17, 18]
 or  secondary standards from Galadi-Enriquez
 et al. (2000) [on Dec. 20.] or both [Dec. 9],
  taken before or after our observations.
 For all standard star observations, atmospheric conditions were photometric.
  The calibration photometry, obtained from
  large (r=2.1\,arcsec) circular apertures, is used to 
 transform our instrumental magnitudes (b, r)
 into the Johnson-Cousins standard system (B,R).
 We estimate the colour terms (see Eqns.~\ref{eqn1a} and ~\ref{eqn1b}.)
 using 20 calibration stars from Dec. 9 with  standard 
  colours (B-R) in the range $\sim$ 0.8 to 1.6\,mag. 
 While we do not have enough calibration stars to determine
 accurate colour terms  for each night separately, CMa observations were
 always performed under photometrically clear conditions, and so
 we use the  colour terms  from Dec. 9 for each of the observations. 
 For a determination of the photometric zero-point offset(s), we 
 have  4 to 20  calibration stars each night, except for Dec. 15 \& 16., 
 for which the calibration from Dec. 18 is used. 
 With the instrumental fluxes already scaled to photons per second,
the transformation equations for Dec. 18, for example, are 
\begin{equation}\label{eqn1a}
 B  = - b + 24.86 (\pm 0.01) + 0.17(\pm 0.01)\,(B-R)
\end{equation}
\begin{equation}\label{eqn1b}
R  = - r + 24.48(\pm 0.04) - 0.07(\pm 0.03)\,(B-R)
\end{equation}

where the instrumental magnitudes (b,r) are corrected for
 atmospheric extinction. We adopted the atmospheric extinction co-efficients
 given in the ESO/WFI web-pages; those are 
 0.24 and 0.09 for the B- and R-bands respectively.
 The error values given  in Eqns.~\ref{eqn1a} and ~\ref{eqn1b} are the 
 rms uncertainties.

 The survey photometry is filtered using quality-control parameters
 computed by ALLSTAR in DAOPHOT, retaining only stars with acceptable
 CHI and SHARP parameters
   as well as formal magnitude errors 
 $\sigma_{\rm B} <$ 0.15\,mag and $\sigma_{\rm R}<$ 0.15\,mag.
 Galactic coordinates and the number of selected stars per field 
 is given in Table~\ref{survey_params}.

\subsection{Completeness and photometry errors}\label{complt_tst}
 To assess the point-source completeness,
 we performed artificial star tests for a representative
 set of six fields. Specifically, we added a
  total of several thousand stars, in groups of 1300, in the
  (instrumental) magnitude range 15.5 to 25.5\,mag,  to each image. 
  The fake stars that we 
 added at random locations to these largely uncrowded fields 
 increase the number-density by up to 10\%, hence barely altering
 crowding effects.  Stars injected in the
  B- and R-band images did not have the same locations.
    We then determined the completeness fractions,
 defined  as the ratio of recovered artificial 
 stars to the number of the injected ones, at
  0.75\,mag-wide intervals. We brighten the B- and R-
 fake star magnitudes by 0.42 and 0.13\,mag, respectively,  
 to shift them approximately from 
 instrumental magnitudes to the flux-calibrated magnitude scale.
Also, as we use de-reddened CMDs in this paper, we shift the 
 the magnitude scale of the completeness curves accordingly and 
  denote the shifted B- and R-band
 magnitudes by B$^\prime$ and R$^\prime$  respectively in Fig.~\ref{Fig_cmpltnss}.
The completeness at (l,b) = (241.5$^\circ$,-6$^\circ$) is  above 90\% 
  for B$^\prime$ or R$^\prime$  $\la$ 20.6\,mag (see Fig.~\ref{Fig_cmpltnss}).
 In the  matched B- and R-band
 photometry lists  the completeness is less, and is typically 
  above $\sim$ 80\% at B$^\prime$ $\la$ 20.6\,mag.
  This 80\% completeness limit 
  varies by $\pm$ $\sim$ 0.5\,mag across the area surveyed at
  $|$b$|$$\ga$6$^\circ$.

 The completeness at each pointing is reduced by gaps
 between detector chips, bad columns and pixels that are the same for
 each field. 
 Completeness  at the faint magnitude end varies from field-to-field for a
 number of reasons:  differences in the seeing, stellar crowding,   
  the cumulative effective of  
  saturated stars (together with their stray light and ghost images,
  caused by internal reflections),  cosmic rays,
 bright galaxies, and bright, nearby
 moving objects.
 Based on the database of artificial star photometry, we
 can assess the photometry errors (e.g. see Fig.~\ref{Fig_errs}).
 For example, 68\% of all 
  artificial stars and, by inference, all detected
 stars  with  B$^\prime$$\sim$22.6
 and R$^\prime$$\sim$22\,mag, at $|$b$|$$\ga$ 6\,deg, 
 have a photometry error typically below 0.05\,mag, and smaller still
 at brighter magnitudes.

\subsection{Extinction and extinction correction}\label{extinc}
Since all our target fields are at low Galactic latitudes, 
 the line-of-sight dust extinction (and, to a lesser extent,
  its dispersion) within each field
 is of great  importance to the analysis in this paper. 
 Schlegel, Finkbeiner \& Davies (1998; SFD98) dust maps provide
 extinction estimates for high latitude fields, but 
 are likely to be inaccurate near the Galactic
 mid-plane, particularly at $|$b$|$$<$  10$^\circ$, according to SFD98. 

As we cannot rely on SFD98 corrections alone, we
 perform an empirical foreground correction of the CMDs
  in two steps. For the  first step 
 we take the SFD98 dust maps,
 interpolated on a star-by-star basis.
 Adopting a standard
 Galactic extinction law  (Cardelli, Clayton \& Mathis 1988) 
 we then determine modified de-reddening values based on the
 prescription given by
 Bonifacio, Monai \& Beers (2000; their Eqn.\,1),  
 and apply those to the photometry; the resultant differential
 extinction data is denoted simply 
 as E(B-V)$_{\rm SFD*}$ hereafter for the sake of clarity. 
 However, the colour of the 
 resulting  CMD morphology in different fields shows that
 this cannot be the full reddening correction. Therefore, we
 apply a second correction
 step, enabled by the prominent main sequence feature in the resulting
 CMDs and based on the assumption that the metallicity and age 
 of the stellar population 
 that causes this CMa MS over-density (e.g. see MD05) does not vary much
 from field-to-field (though its distance may vary). 
With this assumption, the de-reddened colour of the main sequence 
 turn-off (MSTO) stars should be the same in each field. Therefore, we
 shift the stars 
 in each CMD along the reddening vector to match 
 the colour of the  MSTO region
\footnote{We  refer to the blue 
  near-vertical/arc-like  over-dense 
  region at B$_0$$\sim$18.5-to-20 \,mag (E.g., see
  Fig~\ref{Fig_Select_CMD_single})} at [l,b = 242.5$^\circ$, -9.0$^\circ$;
 E(B-V)$_{\rm SFD*}$ =0.148\,mag], which is  a relatively 
 low extinction  field  with  a significant CMa signal.
 This shift is of course a shift both in colour and magitude. 
 To obtain each MSTO colour we make a coarse estimate 
 (number-weighted centroid) at
   B$_0$ = 18.5-19\,mag, and then re-compute it
  in a  0.5\,mag-wide colour interval, centred on the coarse estimate, 
  binned at 0.03\,mag intervals.  
 Outlying 
 bins are ignored by setting a bin-count threshold, taken as the median
 bin-count. 
When testing the de-reddening procedure, we identified obvious failures 
through a visual inspection of the CMDs, focussing on the colour of the
 MSTO region.
  As the CMa MSTO is not well  defined in every field, we 
 {\it do not} apply this additional de-reddening step 
 if E(B-V)$_{\rm SFD*}$ $\le$ 0.11\,mag or if the field is outside 
 -14.7$^\circ$ $\le$ b $<$ 4$^\circ$. De-reddened fields 
 near b$\sim$-15$^\circ$ and also at 
 b $\ge$ 4$^\circ$  have no significant CMa content, 
  and therefore can appear too red by
  $<$ 0.1\,mag. However,  as the density of stars near (B-R)$_0$ =1\,mag
  appears to vary smoothly in our CMDs at such Galactic latitudes,
   this effect  is only a minor concern throughout this paper.
 {\it  Magnitudes and colours   de-reddened in the above-mentioned
 fashion are denoted by}
 B$^{\prime}_0$ {\it and}  (B-R)$^{\prime}_0$ {\it respectively in
 this paper}, and by  B$_0$ and  (B-R)$_0$ otherwise.

 The (MSTO) colour shifts resulting from the de-reddening process, detailed
 above,  are also given in
 Table~\ref{survey_params}; those values  tend to grow with increasing
 Galactic longitude, and  tend to diminish away 
 from the Galactic mid-plane. The values    
  suggest that the SFD98 data both over- and under-estimates the
  foreground extinction across our survey area. 
In order to delimit the impact
  of de-reddening errors further, we take an (ad hoc) 
 extinction threshold E(B-V)$_{\rm  SFD*}$=0.30\,mag. 
E(B-V)$_{\rm SFD*}$ data is given in Table~\ref{survey_params} for each field.

 A final issue is the reddening dispersion in each field. The only attempt
 at compensating for it in this paper is through the first 
  de-reddening step, detailed above,
 which we apply on a star-by-star basis. As the actual reddening 
 dispersion is unknown, we can only proceed on the
 assumption that the predominant variation from field-to-field is in
 the  (median) foreground dust extinction, which we have attempted to
 correct for, and that there are no
 significant differences in the 
 residual (foreground)  dispersion from field-to-field.

\section{Control fields}\label{controlfields}
 The overall density and the line-of-sight distribution of the
 outer Galaxy's stellar constituents (halo, thick/thin disk) vary with 
 position on the sky.
 As we do not precisely model how the known (locally) axisymmetric contributions 
 vary in the area  of sky sampled in this paper,  we  try to minimize 
 the dependence of our results on the current generation
  of synthetic MW models.  We  therefore adopt an
  empirical estimate of the MW components' contribution to
 our CMDs. To illustrate how we select such `control fields',
  we plot in  Fig.~\ref{Fig_compare_Control_cmds} 
  a sub-set of CMDs for three fields of similar $|$b$|$, from 
 above and below the Galactic mid-plane.
 It is immediately apparent from this that  the CMa over-density is not present
 in each field, and those without obvious 
 visible evidence of the CMa-over-density are termed control-fields.
 As we have control fields both above and
  below the Galactic mid-plane, we take the simple
  approach of  using  the field at (l,b) =(240$^\circ$,+8$^\circ$) for the
  latitude range  -14$^\circ$ $\le$ b  $\le$ 8$^\circ$, and take  the
  (240$^\circ$,-20$^\circ$) field 
    when b $<$-14$^\circ$.  At b$>$ 8$^\circ$, 
  the field at  (240$^\circ$, +15$^\circ$) is the control field. 
We stress  that the full imaging survey
will provide a larger array of control fields for a better assessment
and usage of them in future analyses.

\section{Colour-magnitude diagram morphology and the stellar
 populations towards CMa}\label{stel_content}

\subsection{CMD morphology and stellar content of a control field}\label{controlCMD_sec}
Fig.~\ref{FigControl_cmds} shows the CMD of a  field that we take as a
 control field from the  current phase of the imaging
  survey, together with the Besancon MW model
 counterpart\footnote{Default Besancon model via the online web-page
 at  bison.obs-besancon.fr/modele/; Robin et al. (2003)}.  
 It was chosen as a  control-like field 
 because  it shows no obvious presence of the old CMa MS reported in
 B05 and MD05,  based on a visual inspection. 
 The Besancon model is meant to be a description of a
  MW galaxy without inhomogeneities (e.g. spiral arms). As such
  it is very useful 
 for exploring what types of stars at what distances and from which MW
 components (halo, disk, spiral arms)
  contribute to different parts of the CMD.

 The CMD in Fig.~\ref{FigControl_cmds} comprises a range of stellar
 populations at widely differing distances.  The most prominent
 feature of the control field and its model counterpart is 
 the so-called blue-edge of MSTO stars. The colour of this edge 
 at B$^\prime_0$ $\la$ 19\,mag is 
 influenced by  metal-rich thick disk stars and probable metal-poor halo 
 MSTO stars. It is $\sim$ 0.1\,mag bluer at fainter magnitudes, where it is 
 dominated by metal-poor, outer-halo MSTO stars (spectral type F).
 We note   
 (1) that apart from possible photometry scatter,  some of the 
 stars at (B-R)$_0$ $<$-0.4 and B$_0$$\ga$ 19\,mag  may  be 
  nearby  ($<$1.5\,kpc)  white dwarf stars. Blue 
 horizontal branch stars in the halo and thick disk would also have  
  such blue colours. 
     (2)  At B$_0$ $\la$ 15 and (B-R)$^\prime_0$ $\la$ 0.6\,mag in the
 Galactic model CMD, we see that one can expect  
 nearby thin/thick disk MS stars within about 2\,kpc. 
 Further, (3) there is the generally  truncated appearance of the stars
 distributed on the red side of the CMD,
  including a red plume of late-type (M) MS
 stars, which can appear tilted in some fields.  As can be seen in  
 Fig.~\ref{FigControl_cmds}(bottom), those stars
 are nearby thin and thick MS  stars within  about 2\,kpc, and
   to a lesser extent, red
 giant branch stars farther out. 

 Fig.~\ref{Fig_compare_Control_cmds}  illustrates
  that the structure of the CMD in the Galactic North and in the South
are  different over the full latitude range at l$\sim$ 240$^\circ$. 
A comparison with Fig.~\ref{FigControl_cmds}(bottom)  shows 
 that the structure  of the  control field CMDs can be
 reproduced using a standard Galactic model such as the Besancon
  model, but below the midplane an extra population is needed to  
 create the CMa over-density, which is attributed to the 
  MS of a distinct stellar system near D$_\odot$=7.5\,kpc in B05 and MD05 (their
 Fig.1); see Sec.~\ref{interp} also.

\subsection{CMD morphology and stellar content of CMa fields}\label{CMa_stellar_content}
  Fig.~\ref{Fig_Select_CMD_single}
  explains what one expects to see in a typical CMa CMD. 
  Comparing the control fields (i.e. b=+15,+8,-20$^\circ$ in
 Fig.~\ref{Fig_cmds1})  and this CMa field,
  we can now discern  more robustly the  same CMa features
  picked out by MD05 for a 
  pointing near the presumed centre of the CMa over-density.
A prominent feature is the old MS (B05; MD05), which has a
  blue, arc-like (possible MS turn-off) region  near B$^\prime_0$ =
 19\,mag. At brighter magnitudes there is 
 a blue plume of young MS stars
 at [B$^\prime_0$,(B-R)$^\prime_0$]  $\sim$
 [-0.2, 15] to [0.5, 18]\,mag. 
  Further, there is the hint of a 
 red-giant branch in many CMDs, that is visible 
 from [B$^\prime_0$,(B-R)$^\prime_0$] $\sim$
 [15.5, 1.5] to [17.5, 1.2], and  it may contain some CMa stars. 

 Fig.~\ref{Fig_cmds1} to ~\ref{Fig_cmds2} show a large representative
 set of 15 tiles from our sparse map and provide a view  of the
 stellar content  in  the CMa stellar over-density.
 These figures cover  a good range in galactic longitude  and
 latitude, and provide a first  qualitative picture  of the
 stellar content of the CMa stellar over-density over a large angular
 area.  We see that the old MS is a very high-contrast feature
 in several of those tiles, 
   e.g. Fig.~\ref{Fig_cmds1}(centre panel) or  Fig.~\ref{Fig_cmds2}
 (top-right panel). The  density of old MS stars
  exhibits a prominent variation with
 latitude, decreasing away from the Galactic mid-plane (Fig.~\ref{Fig_cmds1}).
 Comparing with CMDs along the longitude direction (Fig~\ref{Fig_cmds2}), a distinct elongation 
 in the old MS stellar population is 
 immediately apparent. In contrast, 
  we see that the young MS population 
 is less evident and less populous at greater
    longitudes. From Fig.~\ref{Fig_cmds1} and Fig.~\ref{Fig_cmds2}, 
   it is apparent that the old MS is still present at b=-15$^\circ$,
  while the younger population is still present at b=-10$^\circ$.

The number-density and shape of the
 young MS star population varies among the CMDs.
 Fig.~\ref{Fig_cmds1} shows that there are dense plumes with 
  a broad  B-band spread, e.g. [l,b =
 240.6$^\circ$, -6.9$^\circ$], but there are also
 narrow, dense plumes, e.g. [l,b = 237.4$^\circ$, -8.1$^\circ$]
  (see Fig.~\ref{Fig_cmds2}). 
  The diversity in the shape (B-band width and density) of the young MS
  is reminiscent of
 the Small Magellanic Cloud (e.g. see CMDs in Noel et al. 2005),
  where there has been 
  on-going and bursty  star formation
 over the past few billon years (Harris \& Zaritsky 2004). 
  The blue MS in the CMa CMDs
  may contain some blue straggler stars. 
   It is likely to be dominated by young MS stars, which is how we
  refer to this population for the remainder of this paper. 
 Carraro et al. (2005) report that this young MS is also observed in
 all of their fields, closer to the Galactic mid-plane, and that this  
 suggests it is associated with the MW spiral galaxy. 
 However, this interpretation 
 may be complicated by the observed differences in its
   morphology (density and magnitude width) at 
 different galactic longitudes (Fig.~\ref{Fig_cmds2}).

 This young (CMa) MS star plume should not be confused with the other similarly blue
 swath of foreground ($\la$ 2\,kpc) white dwarfs 
 at fainter magnitudes in most of 
 our CMDs. Exceptions occur close to the Galactic mid-plane, owing to 
 the possibly inaccurate de-reddening of such nearby stars.
 The young (CMa) MS stars are also different from 
 young ($<$2-3\,Gyr)  foreground ($\la$ 4\,kpc) thick/thin disk MS stars 
 that can occur at similar and  brighter magnitudes 
 than the young MS stars [e.g., at B$^\prime_0$ $\la$ 19,
  (B-R)$^\prime_0$$\la$ 0.4\,mag] marked in
 Fig.~\ref{Fig_Select_CMD_single}.
  Such MW disk stars can be seen
 at b=5.0$^\circ$ and b =-5.9$^\circ$ (l=240$^\circ$) in
 Fig.~\ref{Fig_cmds1}, for example.  
 Based on the model CMD  [see Fig.~\ref{FigControl_cmds}(top)],
 those stars are 
 foreground thin/thick disk MS stars at D$_\odot$$\la$2\,kpc.
 That such MS stars are not detected
 in every field may be a consequence  of field-to-field differences in
  the saturation magnitude, which in turn depends on 
 foreground dust extinction,\footnote{Sufficient foreground extinction
  can cause otherwise undetected (saturated) stars 
  to be detected by dimming them.}  and observing (atmospheric) conditions.

\section{Mapping CMa on the sky}\label{ms_counts}

\subsection{Method -- old and young MS}\label{density1a}
 To estimate  the old  MS star-counts
  that are attributable to the CMa over-density
  at each pointing, we apply a simple
 analysis of the de-reddened photometry 
  using CMD extraction boxes. 
 We estimate the number of 
  old CMa MS stars, N$_{\rm CMa MS}$, in the relevant 
 extraction box (see  Fig.~\ref{FigStarCnts_MS_ref_box_in_cmd}) 
  at a given pointing (l,b) as follows:

\begin{equation}
  N_{\rm CMa MS} = \\
  \gamma\,[N_{\rm MS\,\,box}
 -  {N_{\rm control\,\,field; MS box} \over N_{\rm control\,\,field;
  ref. box}} N_{\rm ref. box}], 
\end{equation}

where ${\rm N_{\rm control\,\,field; MS box}}$ is the star-count in the MS 
 extraction box in the relevant
 control-field. N$_{\rm MS\,\,box}$ and 
N$_{\rm ref. box}$ are  the star-counts at a given (l,b) 
in  the CMa old MS box and 
 a reference box respectively. These boxes are outlined
  in  Fig.~\ref{FigStarCnts_MS_ref_box_in_cmd}.
 $\gamma$ is the scale factor needed to
  compensate for the fraction of area (actually star-counts)
 at 16 $<$ B$_{\rm 0}^\prime$ $<$ 20 above the adopted extinction threshold, 
 E(B-V)$_{\rm SFD*}$ =0.30\,mag, in each field. $\gamma$ = 1 for the
 majority ($>$ 90\%) of these fields, and 1-to-2 otherwise.
 A density estimate and its (random) uncertainty
 for each pointing is then obtained through data re-sampling.
 I.e., 100\% of the stars are 
  selected at random, allowing repeated selection,
  and a MS width estimate is recorded. After repeating
 this 100 times, we fit a Gaussian function
 to a histogram of the density estimates. 
 The mean value from the fit is taken as the density estimate
 and the width
  parameter from the fit is taken as an estimate of the associated 
 (random) error.   Both
  of them are recorded in Table~\ref{survey_params}.

We also map the surface density of young MS stars. 
We do this by counting stars in the extraction box overlaid on the young
 MS in Fig.~\ref{FigStarCnts_MS_ref_box_in_cmd}. 
  While there are many fewer young MS stars than old MS stars, 
 any contamination of the former by field stars 
 appears to be negligible, based on either a visual
 inspection of  our CMDs or the Besancon model CMD 
 in the same direction.
 For that reason we do not subtract a background estimate. Each
  density estimate and its 
 error  (Table~\ref{survey_params}) is  determined 
 through data re-sampling.

\subsection{Results: Comparison of young and old MS star-count profiles}\label{analysis1a}

Fig.~\ref{Fig_MS_BP_RC_profiles} shows the estimated number of young
 and old  MS  stars per WFI field against Galactic latitude and longitude. 
These are the  key results:

  The old MS stellar over-density is highly elongated in Galactic
  longitude in the area of sky
  surveyed  [also see  Fig.~\ref{Figcountprofile}  (bottom)].
 Although complicated by low-latitude extinction (and missing sky
  coverage), the position of its peak density
  appears to be located at (l, b) =  ($\ga$240$^\circ$, $\ga$-7$^\circ$),
  based on a  visual inspection.
 In turn, this suggests that  CMa lies predominantly  (e.g. $>$60\%)
 below the Galactic mid-plane, as found in the 2MASS red-giant
  star analysis in Martin et al. (2004a).
 The survey data  constrain the projected
 aspect ratio of the old stellar over-density.
 There is little surface density gradient in longitude across the
  survey area and it is therefore
 quite reasonable to infer that its FWHM$_{\rm l}$ exceeds the survey width,
 i.e. $>$ 27$^\circ$.
In the latitude direction, the angular width estimate
 of the CMa  density profile is complicated by the fact that the
 density maximum can only be bracketed to be at  b= $\la$ -7$^\circ$
in  Fig.~\ref{Fig_MS_BP_RC_profiles}.   
 Taking  the
 turn-over to be at b$\sim$-7$^\circ$, 
  the FWHM$_{\rm b}$ is $\sim$ 6$^\circ$, based on a visual
 inspection, and CMa's
 projected aspect ratio is therefore  $\ga$ 5:1. 
 For the young  MS, taking FWHM$_{\rm l}$  $>$ 20$^\circ$, conservatively,
 and FWHM$_{\rm b}$ $\sim$ 2$^\circ$, would give
 a projected aspect ratio of $>$  10:1.

The young and old MS
  star populations overlap  on the sky, but  the 
 young stars are markedly more localized in latitude. 
 The detection of a 
    compact  distribution of young stars in the longitude direction
  is especially significant because of the essentially uncontaminated sample.

 The over-density  of young MS stars is not as
  extended  in projection as that of the older stellar population [see
 Fig.~\ref{Fig_MS_BP_RC_profiles}], but is possibly more `flattened.'
 The young MS stars exhibit  
 a maximum density at (l, b) $\sim$ (240$^\circ$,  -7$^\circ$),
 with  a drop-off in their counts at 
 l$<$240$^\circ$ and l$>$ 240$^\circ$, as  can  be verified qualitatively
 from a visual inspection of the CMDs in Fig.~\ref{Fig_cmds2}.
 The same basic conclusions regarding those stars can also 
 be drawn from Fig.~\ref{Figcountprofile}, 
 which  shows a sky-map of their number-density.
   Lastly, there are wiggles in the young and old MS
  profiles, but there is the possibility that they result from having a small
 set of control fields, and/or unknown reddening dispersion.

\section{Line-of-sight depth of CMa at (l,b) = (242.5$^\circ$,-9$^\circ$)}\label{los}
MD05 argued that the quite narrow distance range of the CMa stars points
towards a disrupting satellite, rather than a flaring or warping of
the outer Galactic disk. Here, we explore the line-of-sight extent of 
 CMa by estimating the MS width in  the magnitude direction 
  for (l,b) = (242.5$^\circ$,-9$^\circ$), a low-extinction 
 field near the presumed centre of CMa.
  As this estimate is 
 expected to be the same over
 the CMa over-density we compare with the estimated MS width in MD05
 for (l,b) = (240$^\circ$,-8$^\circ$).

\subsection{Estimating the MS width}\label{method_los}
 We  model the B-band star-count 
 distribution, f$_{\rm total}$ $(\rm B^\prime_0)$,  in a given  CMD  
 colour slice as the linear sum of two components, 
  namely a smooth (underlying) distribution
 of stars from the Galactic stellar halo,
  thick and thin  disks, f$_{\rm bkg}$$(\rm B^\prime_0)$, 
  and the CMa MS,  f$_{\rm MS}$$(\rm B^\prime_0)$: 

\begin{equation}\label{eqn2}
\rm f_{\rm total}(\rm B^\prime_0)  = a\, f_{\rm bkg}(\rm B^\prime_0) + A \,
f_{\rm MS}(\rm B^\prime_0),
\end{equation}

where f$_{\rm  MS}$$(\rm B^\prime_0)$ = [1./($\sqrt{2\pi} \sigma_{\rm
  B_{\rm 0}^\prime}$)]\,EXP[-(B$^\prime_0$ - $\bar{\rm
  B^\prime_0}$)$^2$/(2$\sigma_{\rm B^\prime_0}^2$)], and the
  coefficients (a, A)  are scale factors. We describe the
 background distribution of stars, f$_{\rm bkg}$$(\rm
  B^\prime_0)$, as a second order polynomial, as detailed below, 
 and allow only its amplitude  to vary.

In order to have a compact  B-band
distribution of CMa stars, while having a 
wide  colour slice, 1.0$<$ (B-R)$^\prime_0$  $<$ 1.1\footnote{The colour
 interval (mean colour and width) is  compromise taken between 
   incompleteness at faint magnitudes (and red colours) and the
   rising steepness of the MS in the CMD at blue colours.}  , for signal-to-noise reasons, we 
consider the distribution (B$^\prime_{0,*}$ = ) B$^\prime_0$
-3.5[(B-R)$^\prime_0$ - 1.05],   instead of B$^\prime_0$ alone. The 
  slope is chosen to closely match the slope of the old MS in our CMDs. 
  The method is illustrated in Fig.~\ref{FigMSwidth} for  (l,b) =
  (242.5$^\circ$, -9.$^\circ$), a field near the presumed centre of CMa,   and
  the reference field (l,b) =  (240$^\circ$, +8$^\circ$).  
 The analysis differs from MD05, who had no control field, and 
 a fainter magnitude limit, which allowed them to 
 consider a narrower  colour slice further down along the MS.
The functional form of  f$_{\rm bkg}$ is given in
 Fig.~\ref{FigMSwidth} (top-left panel), with its 
 amplitude scaled to match f$_{\rm MS}$ at  B$^\prime_{0,*}$ = 18-19\,mag,
  binned at 0.25\,mag intervals.   After subtracting  f$_{\rm bkg}$,
  f$_{\rm MS}$ is fitted. The error
 introduced by the binning and shot noise is assessed 
  through data re-sampling.
For the observed MS width, we obtain  $<$$\sigma_{\rm
 B^\prime_{0,*}}$$>$ =   0.49 $\pm$0.05\,mag.

In the above construction, we have assumed a 
symmetrical distribution of star-counts
N(B$^\prime_{0,*}$), which appears to be a reasonable first approximation
  based on a visual inspection of Fig.~\ref{FigMSwidth}(bottom-left).
 To convert the estimated width of the observed MS, 
 $<$ $\sigma^\prime_{\rm B_{0, *}}$$>$, into a
physical line-of-sight (l.o.s.) depth, 
 we firstly subtract in quadrature  the contribution from 
 the photometric uncertainty,
 and the intrinsic width, based on an (assumed) stellar population 
 with a zero  distance range, thereby obtaining  $\sigma_m$. 
  In principle, a reddening dispersion should also
 be subtracted, but our only available way of attempting this is 
 already implemented through the star-by-star de-reddening
 detailed in Sec.~\ref{extinc}. 
 We then transfer $\sigma_m$ to a  physical depth estimate by making
 use of   m - M = -5 + 5\,Log$_{10}$ D$_\odot$, where
 D$_\odot$ (in parsecs) is the 
 heliocentric distance of a star of magnitude m. The determination of
 $\sigma_{\rm los}$ is complicated by  the fact that the adopted
 f$_{\rm MS}$ may not be completely accurate, as remarked in
 L\'opez-Corredoira (2006). A good
 first approximation for the physical depth, in  terms of the
 line-of-sight 1$\sigma$ `radius' or depth, is this:

\begin{equation}\label{eqn1}
\rm \sigma_{\rm los} \approx D_\odot \, [10^{\sigma_{\rm m}/5} - 1]
\end{equation}

 where D$_\odot$ is the distance to the barycentre of the CMa
  over-density along the line-of-sight. It is taken to be 7.5$\pm$1\,kpc, 
  a mid-way estimate between 
 values in the literature [7.2$\pm$1\,kpc, B06;  5-8\,kpc, MD05].

\subsection{An upper limit on the  line-of-sight (l.o.s.) depth of CMa}\label{upp_lim}
To estimate the depth of the old and young  
 CMa over-densities, we  compare the width of its MS to that of a
 reference population.  We now consider the old
 MS population and return to the young MS later below.
 For the old over-density, the reference is a 
 simple stellar population with essentially no depth 
  (i.e. $\sigma_{\rm los, ref}$/D$_{\odot, \rm ref}$ $<$$<$ 1). 
 Thus, by construction, we
 do not account for a likely spread in 
 the chemical composition of the (old) CMa stellar population
 and can provide therefore
  only an  upper limit estimate of its line-of-sight extent.

 For the old reference stellar population, we 
 use photometry of the low-concentration,  high-latitude
 (b=-47.7$^\circ$)  Galactic globular cluster
  Pal 12 from  Martinez-Delgado et al. (2002; see
  Fig.~\ref{Fig_Pal12}, left).  Pal 12  was observed
 through nominally the same filters and has a limiting magnitude similar to the
 CMa data. Taking E(B-V)$_{\rm SFD98}$=0.037\,mag
 and a standard Galactic reddening law (see
 Sec.~\ref{extinc}), we apply a single de-reddening value to the Pal 12
 photometry.  
 Note that we chose the Pal 12 comparison for its matching filter and
  S/N; it does have a slightly different metallicity
  [[Fe/H](Pal 12)= -0.94\,dex, Harris 1996]  than the old CMa MS
  ([Fe/H] $\sim$  -0.7\,dex, B04). 
 As it is a very sparse cluster, we use the inner
  region (r$<$ 1.5\,arcmin), which reduces the
   field-star contamination.   We then apply essentially 
 the same procedure to  the Pal 12 photometry 
(Fig.~\ref{Fig_Pal12}, right) as in the CMa analysis, except that we
 use a slightly wider colour range to have a well-defined MS.
 We obtain $\sigma_{\rm MS,  Pal\,\, 12}$ = 0.16$\pm$0.02\,mag  through
 data re-sampling.

For the young over-density, we  adopt two single-metallicity reference
 stellar populations spanning a wide age range as
 found for the young MS (Carraro et al. 2006, B05, MD05).
 Specifically, we 
 consider two metal-rich (Z=0.006) synthetic
 populations\footnote{Synthetic photometry was obtained using the
 IAC-STAR code (Aparicio \& Gallart 2004).}, namely
 1-13\,Gyr and 0.1-13\,Gyr, at the adopted D$_\odot$=7.5$\pm$1\,kpc,  
 without binaries,  with an adopted constant star formation rate and a zero
 distance range.

\subsubsection{Result -- Depth of the old  MS}\label{res_los}
 We compute the depth $\sigma_{\rm m}$ in magnitudes as follows:
 [0.49$^2$  - 0.16$^2$]$^{0.5}$ = 0.46\,mag ($\pm$ 0.05),
 where the (squared)  values in 
  parentheses are for observed MS width  (see Sec.~\ref{method_los})
 and an estimate of the intrinsic width of a simple stellar population
 (see Sec.~\ref{upp_lim} above). The photometry error is
 negligible based on  Fig.~\ref{Fig_errs}.
 As this is an upper limit,  we have  $\sigma_{\rm m}$$<$  0.46\,mag. 
 Finally, using equation~\ref{eqn1},
 we obtain $\sigma_{\rm los}$  $<$ 1.8 $\pm$  0.2 $\pm$ 0.2  \,kpc, 
 where the errors stem from the uncertainty estimate for $\sigma_{\rm m}$ and
  the adopted distance uncertainty, respectively.  
 Alternatively,  FWHM$_{\rm los}$ $\la$ 4\,kpc.

 As a check, we use data given in  MD05 (their Table 1). 
 Subtracting their tabulated (random) errors, in quadrature, from the 
 total (observed) MS width, we  obtain $\sigma_{\rm m}$ = 0.45\,mag.
 Applying Eqn.~\ref{eqn1}, we have $\sigma_{\rm los}$ = 1.8\,kpc, 
 which is in  excellent agreement with the estimate in the present paper.

\subsubsection{Result -- Depth of the young MS}
We now attempt a first estimate of the l.o.s. depth of the young MS, 
 in a similar way to the old CMa MS analysis. As 
illustrated in Fig.~\ref{Fig_youngMSwidth}, we
 determine the B-band width of the young MS in 
 the  colour range 0.2 $<$ (B-R)$_0$ $<$  0.5\,mag.
  For  the 0.1-13\,Gyr and 1-13\,Gyr synthetic populations, the estimated
 B-band width is  0.37$\pm$  0.17\,mag
 and  0.51 $\pm$ 0.05\,mag respectively. 
 This suggests that there is a 
  dependence on the age of the youngest synthetic stars, and
  a different metallicity may alter this further.
  The  B-band  width of the observed young MS is 0.38$\pm$0.17\,mag.  
 To estimate a preliminary upper limit on the
 l.o.s. depth, we consider the observed MS width plus its uncertaintly
  (i.e. 0.38 + 0.17\,mag), and subtract the width of the
 synthetic MS in quadrature. This gives $\sigma_{\rm m}$ = 0.39\,mag
  and 0.17\,mag
  for the 0.1-13\,Gyr and 1-13\,Gyr populations respectively. Using
 Eqn.~\ref{eqn1} we obtain $\sigma_{\rm los}$ $\la$ 1.5\,kpc and
 0.6\,kpc respectively. 
 Considering that a binary fraction  and a metallicity spread 
 have been omitted, which would decrease the estimated physical depth further, 
 a strong implication
 is that the l.o.s. depth of the young over-density is  $\sigma_{\rm
 los}$ $\la$1.5\,kpc.

\section{Discussion}\label{discussion}
\subsection{Comparison with previous work}
We  have obtained and analyzed a set of 59 CMDs towards the CMa over-density and
overall we have directly presented a representative sub-set of 18.
  Towards CMa, these CMDs reveal a prominent 
 over-density of old and young MS stars at a defined distance range.
 In  Fig.~\ref{FigMSwidth}(bottom-right),  
  one can see that the number-density of old MS stars over the
  expected MW background
 in the colour-magnitude plane is substantial, more than a factor of two 
  for most ($>$ 70\%) of the old MS width. 
 The density profile of old MS stars 
  shown in Fig.~\ref{Fig_MS_BP_RC_profiles}
 provides  unambiguous evidence for a stellar  over-density 
 towards Canis Major.
  This  validates the  2MASS red-giant--based
  detection  of an overdensity in this direction by 
  Martin et al. (2004a). We find CMa to be substantially   
 more elongated in longitude than what Martin et al. 
 found, in general agreement with B06 (see also  Rocha-Pinto et
  al. 2006), whose work was also based on red giants from the 2MASS catalog.

We also presented a  first extended map that traces the 
 surface density map of the young  MS
 stars towards the CMa stellar over-density
 (Fig.~\ref{Fig_MS_BP_RC_profiles}). While a kinematical
 link between the old and young MS populations is lacking, the map  
 shows  they are roughly co-spatial on the sky and that 
 the younger of the two stellar populations is less spatially extended
 in both longitude and latitude.

  Fig.~\ref{Fig_MS_BP_RC_profiles}(right) shows a comparison of 
  the young and old MS profiles together with 
 the red clump surface density profile at b$\sim$-8$^\circ$
  from B06 (estimated from their Fig. 9).
From Table 2 in B06  the  density of red clump stars
 reaches a maximum near l=244$^\circ$ 
  which appears to be compatible with the possible peak in 
  the young MS density profile. No such peak is obvious in the old
  MS data, which are consistent with a near-flat profile in longitude 
 across the  entire  survey width. Qualitatively, the old MS profile  
  appears to be  compatible with the result of Rocha-Pinto et al. (2006): 
 a  large, low-latitude stellar body whose surface density 
  extends beyond l$>$ 270$^\circ$. In this scheme,
  the `original' CMa over-density  (Martin et al. 2004a, B04, MD05)
 would be an outlying part of a larger
  over-density. 
 Overall, it is noteworthy that we detect an over-density of both
  young and old MS stars in the same direction as  the red-clump
  star-count profile given in B06 (their Fig.9). Yet, 
 it is unclear what  actual relationship, if any, exists between them.
  Further study is required to place these issues
  on a firm statistical footing.

We placed an {\it upper} limit of $\sigma_{\rm los}$$<$1.8$\pm$
 0.3\,kpc (FWHM$_{\rm los}$ $\la$ 4\,kpc) on the
 line-of-sight  depth  
 of the (old) CMa stellar over-density and found that it is 
 highly elongated in projection ($\Delta$l:$\Delta$b $\ga$ 5:1). 
 A comparison with  
 the nearest known dwarf galaxy (Sgr), whose RR Lyrae star--based
 l.o.s. FWHM estimate is 5.3\,kpc, corresponding to $\sigma_{\rm los}$$\sim$2.3\,kpc (deduced from (m-M)$_0$ 
data in Cserenjes et al. 2000) only suggests that 
  the CMa depth of a few kilo-parcsecs estimated in this paper 
 is not necessarily incompatible with an origin as a dwarf
galaxy.  We also found that 
 the  depth of the young MS
  population, $\sigma_{\rm los}$  $<$ 1.5\,kpc, is apparently smaller than
 that of  the older stellar population.
   The overall picture portrayed by the data 
 in this paper is one of a young stellar population that is less extended,
 both in terms of 
 its line-of-sight depth and  angular size (both in longitude and latitude),
 than the older population.
 In Sec.~\ref{interp} below,
  we scruntinize this briefly with regard to different interpretations of the
 CMa over-density.

\subsection{On the different interpretations of the CMa over-density}\label{interp}

 As outlined in the introduction, there are three broad classes for 
 the different interpretations of the CMa over-density,
 namely  (a) it is a predominantly old (several
Gigayear-old)  dwarf galaxy  that may be partially 
 disrupted (Bellazzini et al. 2004, 2006; MD05);
 (b) it can be explained by a 
line-of-sight  crossing a warped, but locally axisymmetric
 (i.e. sub-structure--free) outer Galactic disk 
 (Momany et al. 2004, 2006; L\'opez-Corredoira 2006); 
 and (c) the over-densities in the young and old MS stars arise from 
 out-of-plane  MW spiral arms and the Local Arm respectively 
  (Carraro et al. 2005, Moitinho et al 2006).  We discuss these in turn.

(a) If the strongly elongated old over-density that we have mapped is
 the central part of  a dwarf galaxy, could it still be partially bound?
 In  a disrupting satellite, stars are lost from the
   gravitationally bound  main body  and carried in to 
 tidal tails, one leading the 
 satellite and one trailing it.   If a satellite is on
   a   near-circular, low-latitude co-planar orbit, as the CMa dwarf
 would have to be (e.g. Pe\~narrubia et al. 2005), 
 the predominant tidal forces are always in
  the direction towards the Galactic centre. This  leads to 
  perturbations over the entire orbital period,
  allowing stars to  continually 
  escape, unlike eccentric
   orbits, as evidenced by the  modeled disruption of the globular
   cluster  Pal 5  in Dehnen et al. (2004).
  But in such tidal disruption events (e.g.  Piatek \& Pryor 1995,
 Dehnen et al. 2004, Pe\~narrubia et al. 2005), the still bound 
 portion of the stars does not become highly flattened.
 The tidal debris, however, would be wrapped around the Galactic centre, 
 as  modelled by Pe\~narrubia et
 al. (2005) and Martin et al. (2005).
 Based on these generic simulation results, we infer from the strong
   elongation of $\ga$ 5:1 in  longitude that CMa has been undergoing 
 recent tidal disruption and is still localized, but no longer
 gravitationally bound.

 Support for the dwarf galaxy hypothesis has come from the 
  proper motion analysis in Dinescu et al. (2005), which is based on 
 young MS stars in a small area (0.25\,deg$^2$) near the suspected
  centre of the CMa over-density. It reveals a 
 large  (7$\sigma$) deviation from the expectation for stars belonging
 to the warped part of the  Galactic disk,  a motion that fits the
  pre-existing  dynamical model (Pe\~narrubia et  al. 2005), that is
  itself tied only to known parts of the Monoceros 
   stream.\footnote{We note that 
 modeling of the progenitor of the Monoceros stream is discussed in 
Martinez-Delgado et al. (2005b), and shows
a comparison of the N-body Monoceros model with the integrated orbit
  of CMa.}
 This argument of course hinges on the 
   assumption that the young tracer stars 
 are in fact associated with the predominantly older, larger and more
 complex CMa stellar over-density, an issue that is debated  in Carraro
 et al. (2005).  Simulations of dwarf galaxies in dynamical equilibrium 
 suggest that young stars may well be kinematically colder than 
  older populations (McConnachie et al. 2006). 
 However, to firmly establish a link between the young and old CMa
  stellar populations or the absence of it, 
   kinematical (and chemical) study  of both would be essential.

There is one aspect of the projected old and young MS over-density
distributions that is not easily reconciled with the idea of both
arising from a recently disrupted satellite: the large extent of the
young population in longitude, given its more compact distribution in
latitude. If the young stars were at the centre of the putative
precursor  dwarf galaxy (as seen in e.g. Phoenix, Martinez-Delgado et
al. 1999 and Fornax, Stetson et al. 1998),  then they should have been
disrupted last, and hence the least `stretched out' sub-population.
Only yet more extensive imaging (in area coverage) can resolve this issue.

(b) Could the old CMa stellar over-density simply be a consequence of viewing
  the warped outer stellar disk (Momany et al. 2004, 2006) nearly edge-on ?   
 The 2MASS red-giant    star surface density map   
 of Fig. 9 (bottom-panel)  in Momany et al. (2006)  appears  
 to be  broadly compatible with the flat or possibly gradual density 
 variation across our survey width from l=231$^\circ$to 258$^\circ$.
  The angular density distribution of CMa that we have mapped appears,
  taken by itself, to be qualitatively consistent with a
 warped Galactic disk (Momany et al. 2006). 
 From their analysis (their Sec. 4.2), one would infer however 
 that at D$_\odot$=7.3\,kpc star-counts at b=+8$^\circ$
 should match those at b$\sim$ -14$^\circ$. This  is not reflected
 in our data, as is evident 
 from a comparison of the CMDs for b=+8$^\circ$ and -15$^\circ$ in
 our Fig.~\ref{Fig_compare_Control_cmds}.
Rather,
 an additional stellar population is present at l = 230$^\circ$ to
 260$^\circ$, extending  up to 2\,kpc  below the Galactic
 mid-plane.
 
 More importantly, our line-of-sight depth estimate can
 shed  new light on this matter.
 In  Fig.~\ref{warp_check}  we compare the l.o.s. distribution
 predicted by the model in L\'opez-Corredoira et al. (2002; LC02) and 
  Yusifov (2004) to the  one observed here.
  This figure shows (i) that the maximum reached by the 
  old stellar 
 over-density is  significantly farther  [$\Delta$(m-M)$\sim$4.8\,mag
  or $\sim$ 10.5$\sigma_{\rm old\,\,MS}$]  from the maximum expected for a
   locally axisymmetric, warped and flared MW disk, 
  based on LC02 (without spiral arms or other sub-structure). 
 The distinctiveness of the old stellar over-density is still evident if we 
  compare with the pulsar-based Yusifov (2004) model, which yields
  [$\Delta$(m-M)$\sim$2.8\,mag
  or $\sim$6 $\sigma_{\rm old\,\,MS}$].
 In the  Yusifov (2004) model, the density profile has a more gradual
 drop-off with galacto-centric distance, and therefore reaches a maximum
 at larger helio-centric distances than the LC02 model.
    Fig.~\ref{warp_check} also shows 
  (ii) that the observed CMa depth is much smaller than expected for a
 line-of-sight intercepting a warped disk.
  We conclude that the existing `smooth'  (i.e. locally axisymmetric)
 descriptions of the warped outer Galactic  disk  are unlikely explanations of 
  the CMa over-density. As the $\sigma_{\rm los}$ discrepancy
 (distance and width) seems
 generic, variations on the existing warp models are also unlikely to
 match all data.

 Fig.~\ref{Fig_warp_cmd} reiterates this point by showing a colour-magnitude
 comparison of an observed CMa field, a control field and a synthetic
 field for a warped and flared MW stellar 
 disk.  The synthetic population is 
4-10\,Gyr-old  with -0.4 $>$ [Z/Z$_\odot$] $>$ -0.5\,dex, and has been
 convolved with the LC02 MW density profile 
  plotted in Fig.~\ref{warp_check}. It is in
fair qualitative agreement with Fig. 7 in B06, who used a
simple stellar population (47Tuc) as the template population.
 Our model CMD is not meant
to be an exact simulation of the MW's contribution to CMa CMDs, 
but rather it serves to illustrate  that `smooth' (locally
 axi-symmetric) components
  of the warped outer Galactic  disk  alone do not  
 account for the CMa over-density.

 In the model MW CMD,
  Fig.~\ref{Fig_warp_cmd}(right), there is a relative increase 
in star-counts at [(B-R)$\sim$0.8\,mag,  19$<$B$<$17]
 that is not observed in the
control field. It might be a  signature of viewing the warped outer
 Galactic disk in projection, and differs
 substantially from the CMD feature known as the CMa stellar
  over-density, which is evident in Fig.~\ref{Fig_warp_cmd}(left).
 
 Lastly, we note a possibly 
 strong low-latitude (b=-3.4$^\circ$) detection in Bragaglia et
 al. (2006) of the  CMa over-density  towards  the Galactic  anti-centre 
  where the Galactic warp is largely absent. This could be related to a
 tidal tail of the (candidate) dwarf galaxy, and appears to disfavour a
  projection effect of the warped outer MW disk.

 (c) There are interpretations of the young MS 
  population that do not invoke a satellite origin, but 
  instead infer 
  that known types of MW sub-structure may have been detected (Carraro
  et al. 2005, 
  Moitinho et al 2006). In this  scenario  the young stellar
  over-density is  a $\la$100\,Myr-old spiral arm population and the old
  overdensities are merely
   a projection effect of looking along a nearby inter-spiral arm
  structure (the Local Arm).
  While we cannot provide firm evidence against the inter-spiral arm
  scenario, it is unclear whether  
 it can be reconciled with the old stellar population, which
 is still detected at b$\sim$-15$^\circ$ in our survey, and 
 is  distributed over a larger area than the young stellar population.
A comparison of the bluest stars in the empirical and synthetic CMDs 
 in Fig.~\ref{Fig_youngMSwidth} suggests that the most recent
 star formation activity could indeed have  occured less than $\sim$
  100\,Myr ago,
  but this depends  on reddening and the choice of metallicity and
  distance.
 Young ($\la$100\,Myr-old) stars are not necessarily incompatible with
  the MD05 study in which the age used  
 for the youngest MS stars is very uncertain, because of the same 
 degeneracy between 
 the distance, the stellar population (age, Z), and foreground reddening.
 The   youngest stars could be 100\,Myr-old
  if the over-density is farther way (e.g. $\sim$ 10 kpc), as needed
  for the Pe\~narrubia et al. (2005) model, which places the progenitor 
of the Monoceros stream at larger distances too. A reliable
distance estimate, based on RR Lyrae stars for example, as well as
 spectroscopic estimates and extinction  information are
 necessary to solve this controversy unequivocably. We note that young
  (e.g. 100\,Myr) stars themselves are not strong evidence
 against a dwarf galaxy: For example, the last episode of star
  formation observed in  
 the Phoenix dSph was $\sim$ 100\,Myr ago (Martinez-Delgado et al. 1999), 
 and similarly in Sextans A (Dolphin et al. 2003), and in NGC 205 
 (Butler \& Martinez-Delgado et al. 2005).

There is also the issue of why the young stellar over-density can be spotted
   10$^\circ$ (up to 1.3\,kpc) below the Galactic  mid-plane (b=0$^\circ$).
  If we are observing 100\,Myr-old members of a MW spiral arm or an
  inter-spiral arm structure, such stars at D$_\odot$=7.5\,kpc
 would  have drifted away from their parent stellar associations by
 a negligible amount (few $\times$ 100\,pc),
 leaving them absent at b$\sim$-10$^\circ$. This suggests that
 the young stars may have unusual kinematics, as was found for 
  previous   detections of  young ($<$ 100\,Myr) 
 stars several kilo-parsecs from the mid-plane (e.g. Rodgers et
  al. 1981; Lance 1988); such young out-of-plane stars
 have been attributed by those authors to possible accretion events,
   based on stellar kinematics.

\section{Summary and Conclusions}\label{conclusions}
We have presented initial results from 
 our imaging survey of the Canis Major stellar over-density, providing
 a large, representative sub-set of colour-magnitude
 diagrams from our survey area, a depth estimation for the young and
 old MS populations, and a first analysis of their surface density distribution.  In particular, our key findings are these:

\begin{itemize}

\item Using `old' MS stars, we can delineate  an
   over-density of MS stars elongated along galactic longitude at a
   distance of D$_\odot$$\sim$7.5\,kpc.
 It coincides with the 
   over-density of 2MASS red-giants discovered in Martin et
 al. (2004a), but our surface density mapping reveals it to be 
  markedly more elongated than initially thought.
  Its projected aspect ratio is probably
  $\ga$ 5:1,  which is consistent with 
  the more recent  2MASS analyses  (B06; Rocha-Pinto et al. 2006).
We also map the angular distribution of the much bluer MS stars (the `young' MS).

\item The distributions of young and old  stars
  are approximately co-spatial in projection,
  but the young MS density profile is markedly more localized, with a 
  shallow maximum near  (l, b) $\sim$ (240$^\circ$, -7$^\circ$).

\item We report the clear detection of young and old MS stars up to 1.3\,kpc
  and 2\,kpc respectively below the Galactic midplane at
  l$\sim$240$^\circ$.

\item We derive an upper limit on the line-of-sight depth of the (old) stellar 
 over-density, by assuming that its stellar
 population is simple (as, e.g. globular cluster Pal 12), and that its MS
 width soley reflects a distance spread. 
 We obtain  $\sigma_{\rm los}$ $<$ 1.8 $\pm$  0.3\,kpc  
 (or $\sim$ 4\,kpc, FWHM$_{\rm los}$)
 at the adopted D$_\odot$ = 7.5 $\pm$ 1\,kpc.
 The young MS stars are consistent with $\sigma_{\rm los}$  $\la$ 1.5\,kpc.

\end{itemize}

We discussed these results in the context of three broad explanations
 put forth in the literature for the CMa over-density. 
(a) We infer from the strong elongation of the
  over-density in longitude and simulations
  in the literature 
  that, if it is a satellite galaxy on a near-circular, 
  low-latitude orbit, it is unlikely to be still gravitationally bound
  at the present  epoch; it would have to be a recently disrupted satellite.
  (b) The distance and line-of-sight depth of the over-density
 is in disagreement with all published `smooth' (or locally
 axisymmetric)  models of the Galactic 
 warp. Those produce a density profile that is 
 markedly more extended along  the line-of-sight and reaches a maximum
  significantly  closer. 

Lastly, of an out-of-plane spiral arm 
  hypothesis for the young MS stars, we note that 
 the presence of young  out-of-plane stars is not uncommon in the MW,
and such stars in the literature
  have been attributed to possible accretion events,
 based on stellar kinematics.

Without detailed modeling the data themselves  are not yet
sufficient to discriminate between an interpretation as 
 sub-structure in the pre-existing 
 warped outer Galactic  disk  or a disrupted satellite.

\acknowledgments
 It is a great pleasure to thank T. Erben for his very generous
 help in using the Bonn-Garching WFI pre-reduction pipeline.
We thank A. Robin for 
 providing comments on the limitations of the current Besancon Galaxy model. 
This work has made use of the IAC-STAR synthetic CMD computation
 code. This code is  maintained by the computer division of the
 Instituto de Astrofísica de Canarias. DMD recognizes the support of 
the Spanish Ministry of Education and Science (Ramon y Cajal contract 
 and research project AYA 2001-3939-C03-01). JP thanks Julio Navarro for
financial support.

\clearpage

\begin{figure*}
This figure is given separately as a JPG file.
\caption{Colour-magnitude diagram (CMD) for a field near the presumed
 centre of the CMa over-density, at (l,b) = (240.5$^\circ$, -6.8$^\circ$). It 
 is labelled to show several 
 different stellar components, namely  the young
 and old CMa  main sequences, and the red plume of late type
 thin and thick disk MS stars. 
The approximate saturation level is marked (dashes). The CMD photometry
 has been de-reddened, as detailed in Sec.~\ref{extinc}.
See Sec.~\ref{controlCMD_sec} for further
 explanations of the CMD content.
    } \label{Fig_Select_CMD_single} 
    \end{figure*}

\clearpage  

\begin{figure*}
\includegraphics[angle=0]{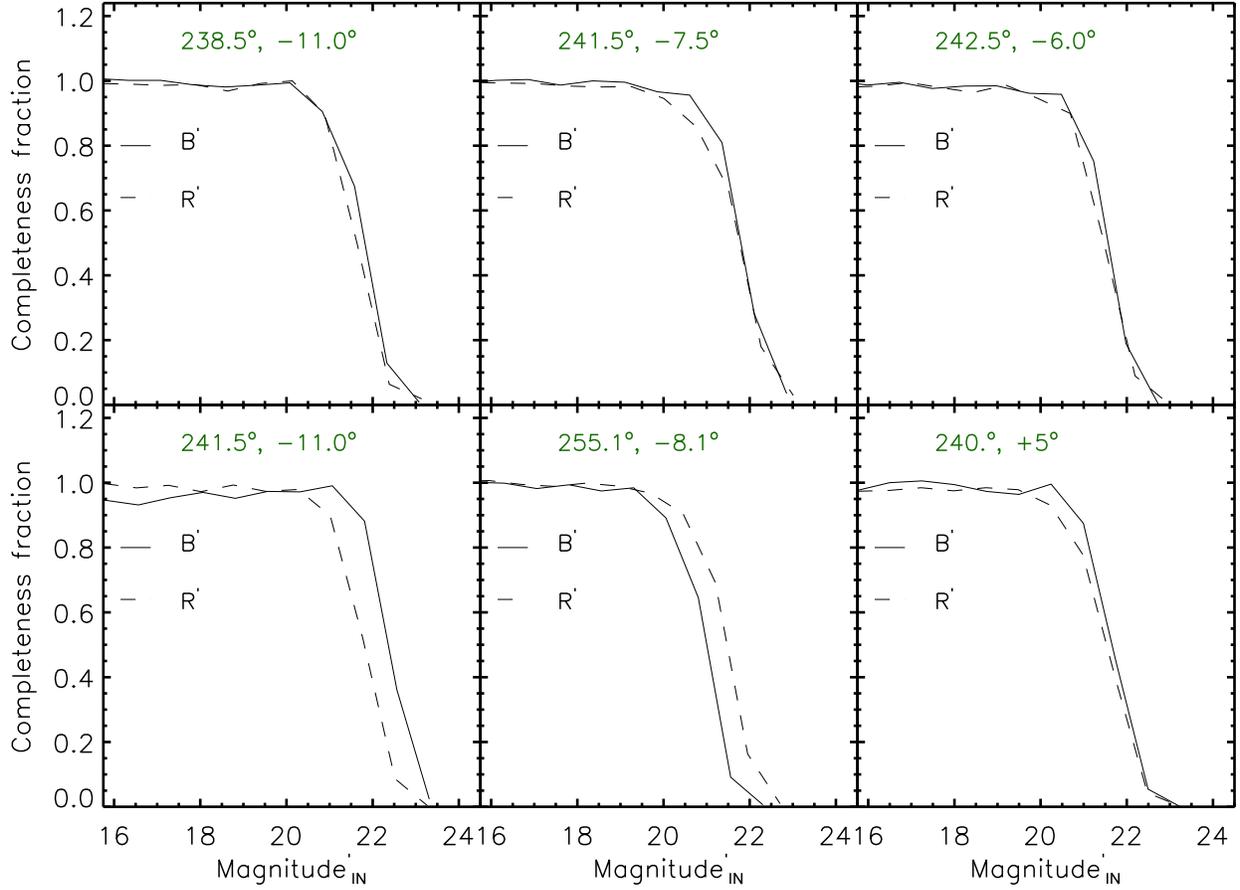}
 \caption{Sample completeness diagrams for the CMa photometry 
 from artificial star tests as a function of the input artificial star
 magnitude. Coupled with Fig.~\ref{Fig_errs}, this is a diagnostic of
 the photometric quality. We have restricted the tests to six fields
 that provide a good coverage of the range of latitudes and longitudes
 of the CMa over-density.
As these tests were performed using instrumental
 magnitudes, we applied an approximate zero-point correction to each
 bandpass to shift the magnitudes to the flux calibrated system.
Also, as we use de-reddened CMDs in this paper, we applied an appropriate 
 shift to the input magnitude scale of each completeness curve, denoted by
 Magnitude$^\prime_{\rm IN}$.
 We refer to the magnitude shifted B- and R-band completeness curves 
 by B$^\prime$ and R$^\prime$  respectively in the panels. 
 See Sec.~\ref{obs_datared} for further details.}
 \label{Fig_cmpltnss}
    \end{figure*}
 
\clearpage

\begin{figure*}
These figures are given separately as a JPG file.
 \caption{Photometry diagnostic, based on artificial star tests.
  In each of  the two groups 
  of panels we provide magnitude scatter plots (upper) [recovered minus
input magnitudes] as a function of the input magnitude, as well as  the magnitude dependence of the
 error curve enclosing $\pm$34\% of the stars (lower panels). 
 They
 indicate that 68\% of injected stars at  B$^\prime$$\sim$22.6\,mag and
 R$^\prime$$\sim$22\,mag  in these fields have a photometry error
 typically below 0.05\,mag, and smaller still at brighter magnitudes.
  For the sake of clarity, we 
 restricted this diagnostic check to a sub-set of six  fields spanning 
  the survey area, with some at similar positions to check
 the  local and global photometric quality.
For the meaning of B$^\prime_{\rm IN}$ and R$^\prime_{\rm IN}$, please
  see the meaning of Magnitude$^\prime_{\rm IN}$ in Fig.~\ref{Fig_cmpltnss}.
} \label{Fig_errs}
    \end{figure*}
 
\clearpage

\begin{figure*}
This figure is given separately as a JPG file.
 \caption{CMD of an example `control field' at (l,b)=(239.9$^\circ$,-20$^\circ$) from our  survey (top)
  and its interpretation in terms of
 a synthetic MW model (bottom), which accounts only for the smooth and
 symmetric components (thin, thick disk, and halo). It shows that
  such fields comprise stars over a range of distances without
  prominent sub-structure along the line-of-sight. There is an
  absence of stars at B$_{\rm 0}$ $\la$ 14.5 in the control field owing
  to detector saturation, and at B$_{\rm 0}$ $\ga$ 22.5 because of the
  magnitude limit.  It can be seen 
 in the Galactic model CMD that fainter MS stars at a given colour are farther away.  The marked density of 
  model stars at  [B$_{\rm 0}$ $>$ 22, (B-R)$_{\rm 0}$ $\sim$ 0.6-0.8\,mag] 
 in the synthetic
  CMD are ancient, metal-poor MS turn-off stars in the outer Galactic halo, but their   number-density may be inaccurate.
} \label{FigControl_cmds}
    \end{figure*}

\clearpage  

\begin{figure*}
\includegraphics[width=16cm,height=15cm,angle=0]{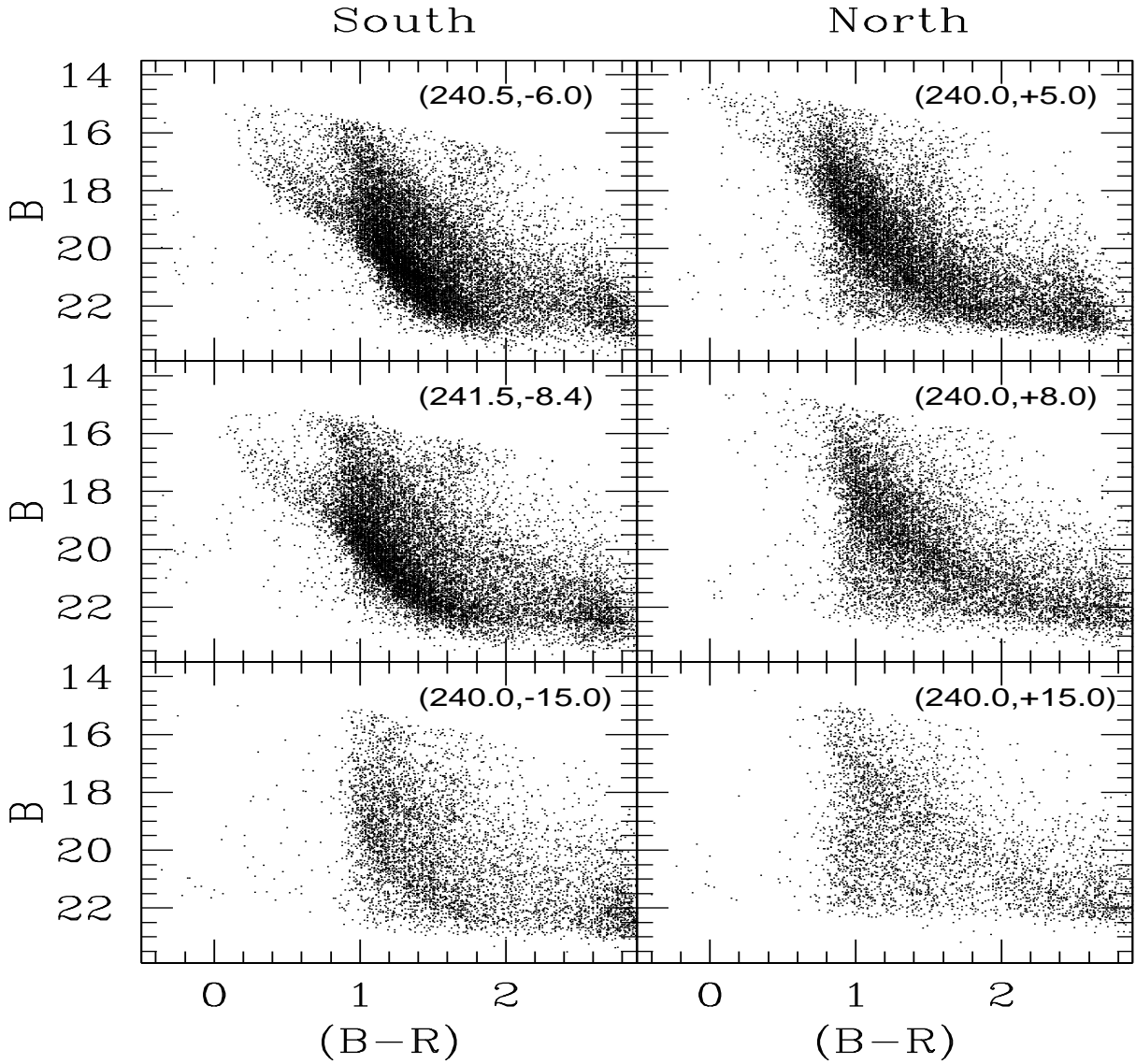}
 \caption{CMD of example control fields (right) at 
  b =(+15$^\circ$,+8$^\circ$,+5$^\circ$) from our  survey (top-to-bottom),
  together with the CMDs for  b =(-15$^\circ$,-8$^\circ$,-5$^\circ$)
  (left). It illustrates that the structure of the CMDs for fields
  above and below the mid-plane differ over the full latitude range at
  l=240$^\circ$.
} \label{Fig_compare_Control_cmds}
    \end{figure*}

\clearpage

\begin{figure*}
This figure is given separately as a JPG file.
 \caption{We 
 plot a sequence of CMDs across the Galactic plane along a fixed longitude
 (l$\sim$240$^\circ$), running from b = +15$^\circ$ to b = -20$^\circ$. 
 The text in each panel refers to the (l,b) position.  
The  axes are labelled [B$_0^\prime$, (B-R)$_0^\prime$] 
because the photometry was de-reddened in two different
 steps (see Sec~\ref{extinc}).  Exceptions are the fields at 
 b=+15$^\circ$,+8$^\circ$,+5$^\circ$ and -20$^\circ$, whose correct axis
 labelling is 
 [B$_0$, (B-R)$_0$],  but is excluded for the sake of clarity. 
 The top-right panel shows photometry error bars (based on
 Fig.~\ref{Fig_errs})  
 at (B-R)$_0^\prime$=0, which are typical for the bulk of these CMDs.
 We also show the de-reddening vector arising from the additional
 (i.e. post-SFD*) de-reddening step, which is described in Sec.~\ref{extinc}.
  The 80\% completeness limit (dashes), is included in two panels,
 illustrating the faint limiting magnitude of the photometry.
} \label{Fig_cmds1}
    \end{figure*}
 
\clearpage  
\begin{figure*}
This figure is given separately as a JPG file.
 \caption{As Fig.~\ref{Fig_cmds1}, but for a sequence parallel to the
 Galactic plane at latitude
 (b$\sim$-8$^\circ$), running from l $\sim$ 234 to l $\sim$
 258$^\circ$.  For other explanations, see Fig.~\ref{Fig_cmds1}.
} \label{Fig_cmds2}
    \end{figure*}

\clearpage

\begin{figure*} 
This figure is given separately as a JPG file.
\caption{Illustrating the extraction boxes used for the young and old MS star
   count estimates. The box to the right of the old MS is a reference
   box of MW stars. The 80\% completeness limit is marked.
See Sec.~\ref{density1a} for more on the extraction boxes. 
} \label{FigStarCnts_MS_ref_box_in_cmd}
    \end{figure*}

\clearpage  
\begin{figure*}
\includegraphics[width=16cm,height=10cm,angle=0]{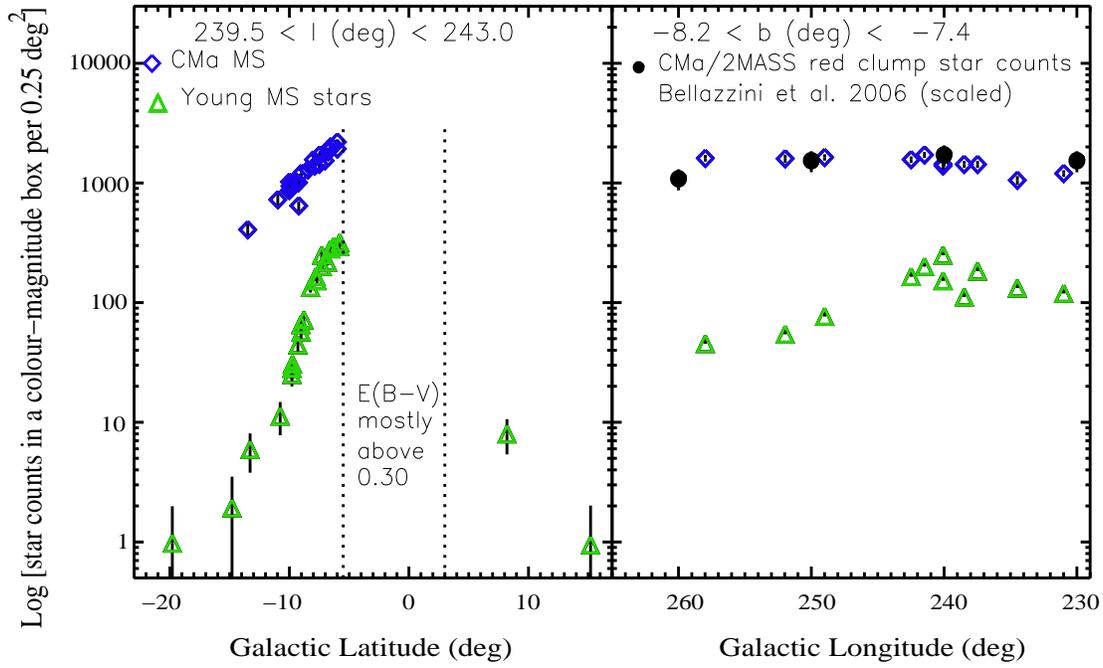} 
 \caption{Mapping the CMa stellar overdensity of young and old CMa  
 MS stars as a function of Galactic latitude (left) and longitude (right).
Only fields with median  E(B-V)$_{\rm SFD*}$  below 0.30\,mag are included. 
The open diamonds represent the density of old MS stars (as estimated
 in Sec.~\ref{ms_counts} and recorded in Table~\ref{survey_params}.); the triangles represent the young
 MS stars. In addition, CMa density estimates from red clump data
  (Bellazzini et al. 2006) are shown in the
 right panel,   based on their Fig. 9, together
  with an assigned  20\% error. The maximum of the Bellazzini et
 al. values has been scaled to match our results.
} \label{Fig_MS_BP_RC_profiles}
    \end{figure*}

\clearpage

  \begin{figure*}
 This figure is given separately as a JPG file.
\caption{ Surface density map of the area of sky sampled,
 including high extinction fields (median E(B-V)$_{\rm SFD*}$ $>$ 0.30\,mag;
 diamonds). For the sake of clarity, the WFI field-of-view is
 over-sized. We note that the pointings at l$>$251$^\circ$
 were originally designed to be control fields.
 The (relative) number-density of old CMa MS stars (bottom) and young MS 
  stars (top) is intensity-coded.
 One sees an over-density of  old MS stars that is most likely centred
 below the Galactic mid-plane (also see 
 Fig.~\ref{FigStarCnts_MS_ref_box_in_cmd}), and exhibits 
 a pronounced elongation in Galactic longitude. The young MS
 population is probably also centred below the mid-plane, but is 
 substantially more localized in the longitude direction.
} \label{Figcountprofile}
    \end{figure*}

\clearpage

\begin{figure*}
\includegraphics[width=16cm,height=12cm,angle=0]{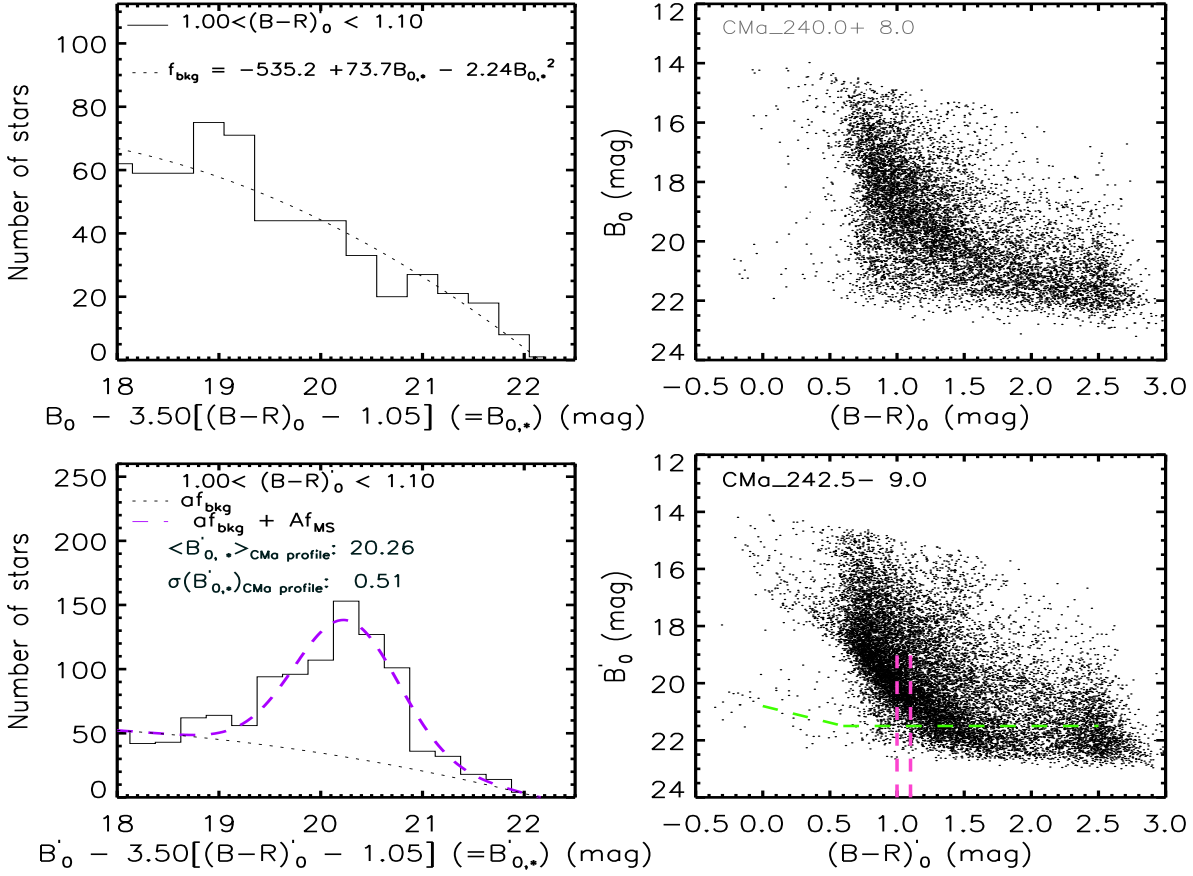} 
 \caption{Estimating the line-of-sight depth of CMa: We show the 
 star-count distribution   (left panels)
 and the associated CMD (right panels), for a reference star-field (top)
 and a field, (l,b) =  (242.5$^\circ$, -9.$^\circ$)  (bottom), near the
 presumed centre of the CMa over-density. 
 The `thickness' of the old MS, reflecting the l.o.s. extent is
 estimated in the colour interval bounded by vertical lines (dashes)  
 in the bottom-right panel. As a guide to the photometric quality,
  the 80\% completeness
 limit is also marked  (horizontal and tilted dashes).
 In the bottom-left panel, 
 the functional form from the top-left panel (short-dashes), together with a 
 Gaussian function is fitted (long-dashes) to the histogram. The data in this
 panel are representative, and correspond to
  only one iteration of data re-sampling. See
 Sec.~\ref{density1a} for  further explanations of the depth
 estimation.   
} \label{FigMSwidth}
    \end{figure*}

\clearpage

\begin{figure*}
\includegraphics[width=16cm,height=8cm,angle=0]{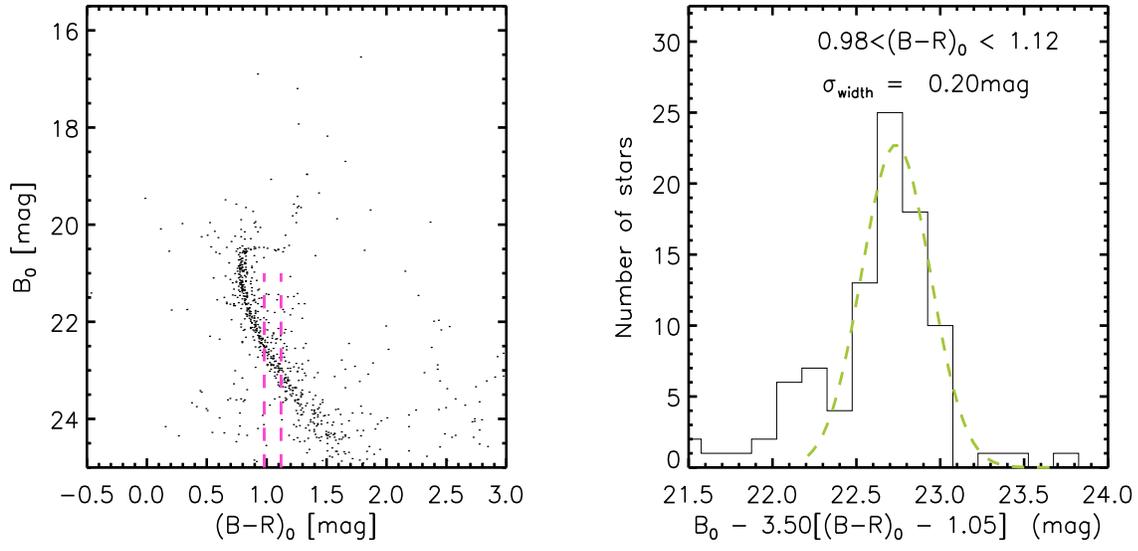} 
 \caption{Estimating the MS width of a reference stellar population, 
 the  globular cluster Pal 12.  We have applied  a similar procedure 
 to that shown in Fig.~\ref{FigMSwidth}. We extracted photometry
 from a narrow colour interval in the CMD (left-panel). We then fitted
  a  Gaussian distribution (dashes, right-panel).  The 1\,$\sigma$
 width data given in the
 right-panel is from one iteration of data re-sampling. 
 For more information, see Sec.~\ref{upp_lim}. 
  } \label{Fig_Pal12}
    \end{figure*}

\clearpage  

\begin{figure*}
\includegraphics[width=16cm,height=14cm,angle=0]{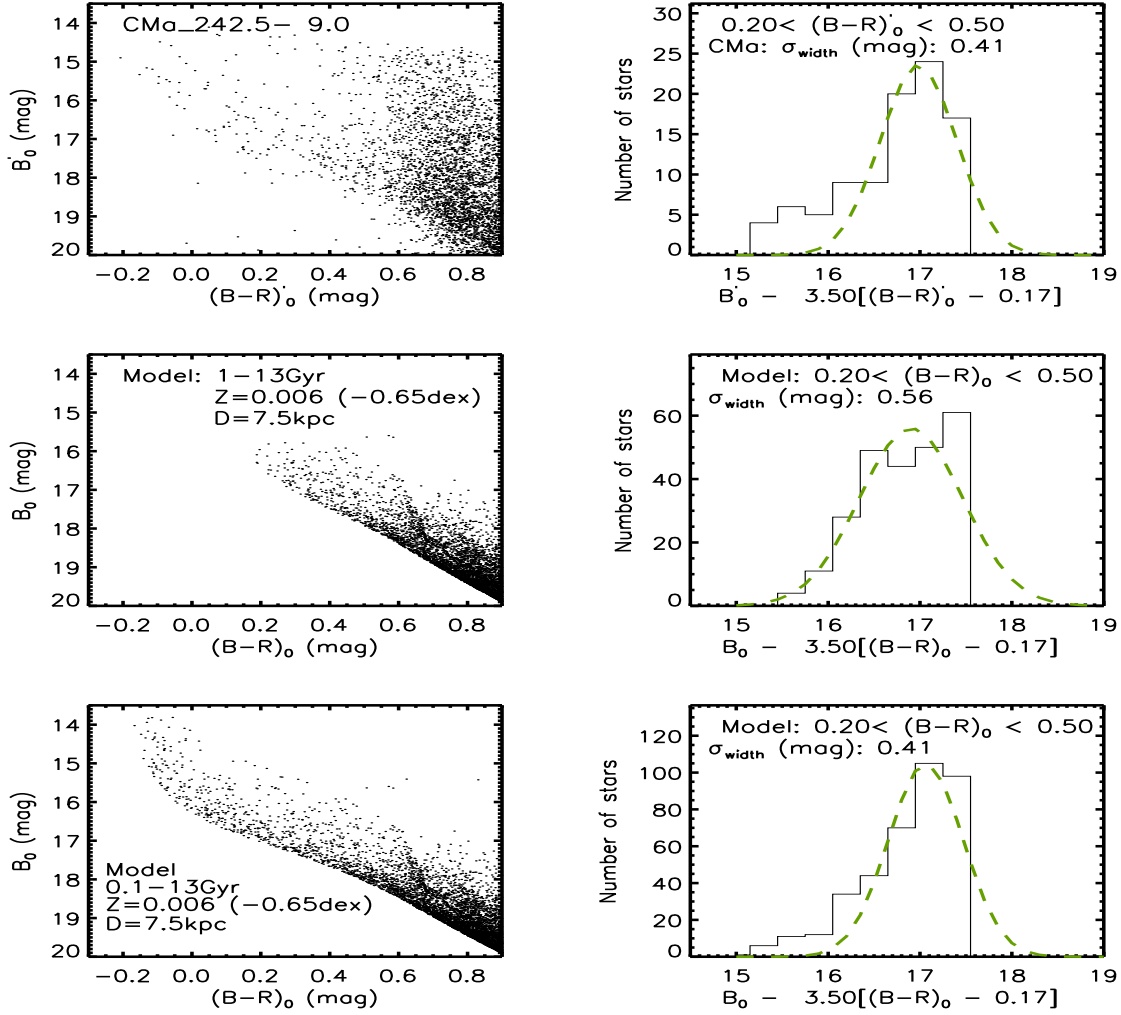} 
 \caption{Demonstrating the depth estimation method for the young MS
   population:  We use same CMD (l,b = 242.5$^\circ$, -9$^\circ$)
 used to estimate the depth of
   the old CMa MS population (see Sec.~\ref{method_los}). We show the young empirical MS (top-left) and a
   corresponding histogram of star counts (top-right) from one
   iteration of data re-sampling, together with a
   fitted Gaussian distribution (dashes). The other panels are
  for the adopted synthetic stellar populations. Additional details on the depth estimation can be found in
   the panels and    in Sec.~\ref{upp_lim}. 
     } \label{Fig_youngMSwidth}
    \end{figure*}

\clearpage

\begin{figure*}
\includegraphics[width=8cm,height=8cm,angle=0]{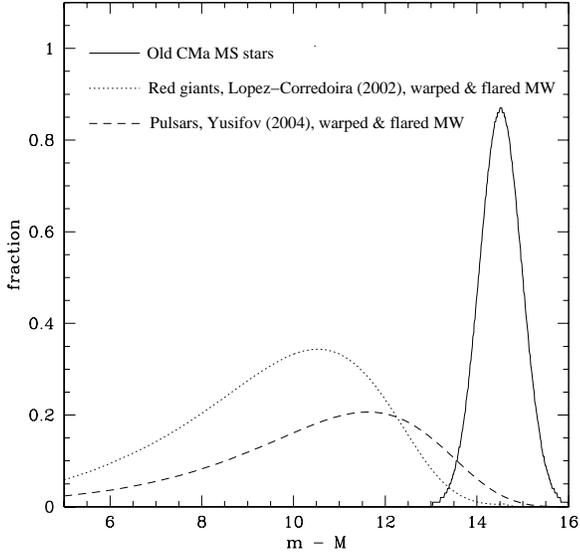} 
\caption{Illustrating that none of the `smooth' (locally axisymmetric)
  MW models for (l,b) =
 (241.5$^\circ$,-7.5$^\circ$) reach a maximum density 
   near the  distance modulus of the old CMa stars (solid).
 The area under each profile has been normalized to unity.
 The MW models are the warped, flared and inhomogeneity-free 
 Galactic disk model from L\'opez-Corredoira et al. (2002) (short
 dashes) and the Yusifov (2004) model (long dashes). Model parameters
 are as given by those authors, except that we curtail the
 amplitude (z$_{\rm w}$) of the warped  L\'opez-Corredoira et al. 
 disk from diverging to infinity by setting z$_{\rm w}$(R$>$14\,kpc) = z$_{\rm
 w}$(R=14\,kpc), where R is galacto-centric distance. 
} \label{warp_check}
    \end{figure*}
\clearpage

\begin{figure*}
This figure is given separately as a JPG file.
 \caption{Illustrating that the CMa stellar
  over-density at (l,b) = (240$^\circ$,-8$^\circ$)  (left) is absent in the  control field at b=+8$^\circ$
 (middle), and is also absent in the (unreddened) Galaxy model CMD for b=-8$^\circ$  (right). 
 The Galactic model has been constructed by convolving  a synthetic
 population (age =4-10\,Gyr, -0.4 $>$[Z/Z$_\odot$]$>$ -0.5\,dex),
  with the LC02 line-of-sight density
 profile of MW stars given in Fig.~\ref{warp_check}.
 Even though it has not been convolved with photometry errors, and
 contains no MW halo stars, a comparison of all three CMDs 
  strongly disfavours the warped MW disk as a way to explain the old stellar
  over-density found in the CMa field.
 As we do not model the red plume of nearby ($\la$ 1.5\,kpc) late-type 
 MS stars, the red boundary is set to (B-R)=2\,mag.
} \label{Fig_warp_cmd}
    \end{figure*}

\clearpage

\begin{deluxetable}{lccclcll}
\tabletypesize{\scriptsize}
\tablecaption{Target information.}
\tablewidth{0pt}
\tablehead{  
\colhead{ID} & \colhead{l} & \colhead{b}   &
\colhead{E(B-V)$_{\rm SFD*}^{\rm a}$ $\pm$ $\sigma$}  & 
\colhead{$\Delta$ (B-R)$_0$}  & 
\colhead{N$_{\rm CMD}$ } & 
\colhead{N$_{\rm Old\,\,MS}$ $\pm$  $\sigma$} & 
\colhead{N$_{\rm Young\,\,MS}$ $\pm$$\sigma$} \cr
\colhead{} & \colhead{} & \colhead{} & \colhead{} &
  \colhead{}  & \colhead{}  & \colhead{} & \colhead{} \cr
  \colhead{} & \colhead{(deg)} & \colhead{(deg)} &
  \colhead{(mag)}    & \colhead{(mag)}    & \colhead{}  & \colhead{} & 
  \colhead{}   \cr 
\colhead{(1)} & \colhead{(2)} & \colhead{(3)} &
  \colhead{(4)}  & \colhead{(5)} & \colhead{(6)}  & \colhead{(7)} 
 & \colhead{(8)}  \cr 
}
\startdata
CMa\_231.0- 8.0        &   230.989        &    -7.995        &     0.235     $\pm$    0.016        &     -0.05        & 11964        &  1236     $\pm$   85        &   117     $\pm$   10
           \cr 
CMa\_233.0-15.0        &   232.976        &   -14.997        &     0.063     $\pm$    0.004        &  \nodata        &  5684        &   -51     $\pm$   70        &     4     $\pm$    2
           \cr 
CMa\_233.0-14.0        &   232.985        &   -13.995        &     0.092     $\pm$    0.009        &     0.00        &  7256        &   183     $\pm$   50        &     7     $\pm$    2
           \cr 
CMa\_234.5- 8.0        &   234.482        &    -8.001        &     0.220     $\pm$    0.009        &     -0.10        & 14641        &  1038     $\pm$   77        &   133     $\pm$   10
           \cr 
CMa\_235.0-10.0        &   234.984        &    -9.996        &     0.188     $\pm$    0.016        &     -0.03        & 10412        &   699     $\pm$   80        &    36     $\pm$    6
           \cr 
CMa\_235.5-11.5        &   235.484        &   -11.504        &     0.135     $\pm$    0.011        &     -0.06        &  9221        &   585     $\pm$   66        &    12     $\pm$    3
           \cr 
CMa\_235.5- 5.0        &   235.528        &    -5.015        &     0.412     $\pm$    0.034        &    \nodata        & 13195        & \nodata        &   \nodata
           \cr 
CMa\_236.5- 3.5        &   236.465        &    -3.505        &     0.740     $\pm$    0.081        &     \nodata        & 11933        & \nodata        &     \nodata
           \cr 
CMa\_237.0-12.5        &   236.983        &   -12.494        &     0.118     $\pm$    0.004        &     -0.10        &  8193        &   487     $\pm$   67        &    13     $\pm$    3
           \cr 
CMa\_237.5- 9.0        &   237.478        &    -8.993        &     0.219     $\pm$    0.032        &     -0.07        & 12360        &  1138     $\pm$   75        &    86     $\pm$    8
           \cr 
CMa\_237.5- 8.0        &   237.480        &    -7.995        &     0.219     $\pm$    0.018        &     -0.02        & 15668        &  1421    $\pm$   96        &   182     $\pm$   15
           \cr 
CMa\_237.5- 6.0        &   237.483        &    -5.997        &     0.305     $\pm$    0.022        &      \nodata        & 19815        & \nodata         &   \nodata
           \cr 
CMa\_237.5- 7.0        &   237.486        &    -7.020        &     0.287     $\pm$    0.023        &     -0.00        & 16331        &  1804     $\pm$  105        &   185     $\pm$   14
           \cr 
CMa\_238.5- 5.0        &   238.476        &    -4.993        &     0.489     $\pm$    0.050        &     \nodata       & 15497        & \nodata         &     \nodata
           \cr 
CMa\_238.5-11.0        &   238.478        &   -10.976        &     0.141     $\pm$    0.007        &     -0.09        & 10235        &   813     $\pm$   65        &    12     $\pm$    3
           \cr 
CMa\_238.5- 6.5        &   238.481        &    -6.500        &     0.277     $\pm$    0.030        &     -0.08        & 13545        &  1507     $\pm$  100        &   186     $\pm$   13
           \cr 
CMa\_238.5- 7.5        &   238.484        &    -7.468        &     0.298     $\pm$    0.021        &     -0.08        & 15544        &  1456     $\pm$  114        &   111     $\pm$   11
           \cr 
CMa\_239.7- 9.2        &   239.680        &    -9.244        &     0.143     $\pm$    0.004        &     -0.00        & 12876        &   993     $\pm$   88        &    66     $\pm$    7
           \cr 
CMa\_239.7-10.0        &   239.681        &   -10.001        &     0.127     $\pm$    0.007        &      0.00        & 12436        &   867     $\pm$   74        &    27     $\pm$    4
           \cr 
CMa\_239.7- 6.0        &   239.687        &    -5.994        &     0.304     $\pm$    0.021        &     \nodata       & 15882        & \nodata         &  \nodata
           \cr 
CMa\_239.7- 6.8        &   239.691        &    -6.754        &     0.338     $\pm$    0.024        &     \nodata        & 15001        & \nodata         &     \nodata
           \cr 
CMa\_240.0+ 5.0        &   239.981        &     4.991        &     0.152     $\pm$    0.004        &  \nodata        & 14213        &   -25     $\pm$  121        &     8     $\pm$    2
           \cr 
CMa\_240.0-15.0        &   239.981        &   -15.000        &     0.100     $\pm$    0.004        &  \nodata        &  5765        &  -129     $\pm$   77        &     2     $\pm$    1
           \cr 
CMa\_240.0-20.0        &   239.985        &   -19.997        &     0.047     $\pm$    0.002        &  \nodata        &  3169        &     2     $\pm$   51        &     1     $\pm$    1
           \cr 
CMa\_240.0+15.0        &   239.986        &    14.996        &     0.052     $\pm$    0.003        &  \nodata        &  3933        &   -39     $\pm$   39        &     1     $\pm$    1
           \cr 
CMa\_240.0+ 8.0        &   239.988        &     8.000        &     0.109     $\pm$    0.004        &  \nodata        &  9543        &   -25     $\pm$   78        &     8     $\pm$    2
           \cr 
CMa\_240.0- 2.7        &   239.999        &    -2.693        &     0.531     $\pm$    0.004        &      \nodata        &  1781        & \nodata         &     \nodata
           \cr 
CMa\_240.1- 7.9        &   240.078        &    -7.894        &     0.211     $\pm$    0.020        &      0.14        & 13132        &  1374     $\pm$  106        &   150     $\pm$   11
           \cr 
CMa\_240.1- 7.5        &   240.082        &    -7.503        &     0.228     $\pm$    0.026        &     -0.21        & 13557        &  1467     $\pm$  128        &   251     $\pm$   16
           \cr 
CMa\_240.1-13.5        &   240.083        &   -13.491        &     0.134     $\pm$    0.009        &     -0.16        &  5795        &   407     $\pm$   55        &     5     $\pm$    2
           \cr 
CMa\_240.5-10.0        &   240.482        &    -9.995        &     0.123     $\pm$    0.004        &      0.12        & 11144        &  1007     $\pm$   55        &    23     $\pm$    4
           \cr 
CMa\_240.5- 6.0        &   240.485        &    -5.996        &     0.239     $\pm$    0.014        &      0.08        & 21239        &  2193     $\pm$  115        &   318     $\pm$   15
           \cr 
CMa\_240.5- 6.8        &   240.490        &    -6.761        &     0.228     $\pm$    0.015        &      0.06        & 19616        &  1783     $\pm$  124        &   280     $\pm$   15
           \cr 
CMa\_240.5- 9.2        &   240.514        &    -9.225        &     0.125     $\pm$    0.004        &     -0.09        &  9495        &   642     $\pm$   64        &    59     $\pm$    7
           \cr 
CMa\_241.5- 7.5        &   241.480        &    -7.468        &     0.184     $\pm$    0.013        &      0.10        & 18257        &  1698     $\pm$   95        &   201     $\pm$   15
           \cr 
CMa\_241.5-11.0        &   241.482        &   -10.975        &     0.118     $\pm$    0.006        &      0.02        & 11286        &   741     $\pm$   87        &    10     $\pm$    3
           \cr 
CMa\_241.5- 9.5        &   241.482        &    -9.495        &     0.133     $\pm$    0.004        &      0.03        & 14146        &  1018     $\pm$   85        &    44     $\pm$    6
           \cr 
CMa\_241.5- 8.4        &   241.487        &    -8.442        &     0.151     $\pm$    0.005        &      0.05        & 15248        &  1285     $\pm$   87        &   131     $\pm$   10
           \cr 
CMa\_241.5- 6.5        &   241.487        &    -6.499        &     0.251     $\pm$    0.019        &      0.11        & 19328        &  2028     $\pm$  106        &   289     $\pm$   15
           \cr 
CMa\_241.5- 5.0        &   241.488        &    -4.978        &     0.343     $\pm$    0.042        &     \nodata        & 22302        & \nodata        &  \nodata
           \cr 
CMa\_242.5- 8.0        &   242.482        &    -7.990        &     0.160     $\pm$    0.009        &      0.07        & 17637        &  1543     $\pm$   92        &   165     $\pm$   12
           \cr 
CMa\_242.5- 7.0        &   242.483        &    -7.016        &     0.193     $\pm$    0.015        &      0.02        & 18419        &  1583     $\pm$  139        &   216     $\pm$   12
           \cr 
CMa\_242.5- 9.9        &   242.484        &    -9.949        &     0.136     $\pm$    0.017        &     -0.04        & 12451        &   929     $\pm$   74        &    31     $\pm$    5
           \cr 
CMa\_242.5- 9.0        &   242.484        &    -8.994        &     0.129     $\pm$    0.009        &      0.00        & 15397        &  1188     $\pm$   80        &    72     $\pm$    7
           \cr 
CMa\_242.5- 6.0        &   242.486        &    -5.993        &     0.258     $\pm$    0.019        &      0.02        & 19347        &  1943     $\pm$  125        &   296     $\pm$   19
           \cr 
CMa\_243.0- 3.6        &   242.999        &    -3.630        &     0.508     $\pm$    0.021        &  \nodata        &  9434        & \nodata         &   \nodata
           \cr 
CMa\_248.0-14.0        &   247.982        &   -13.993        &     0.164     $\pm$    0.008        &     -0.08        &  6477        &   333     $\pm$   61        &     5     $\pm$    2
           \cr 
CMa\_248.0-15.0        &   247.990        &   -14.999        &     0.168     $\pm$    0.016        &     \nodata  &  5666        &    50     $\pm$   69        &     4     $\pm$    1
           \cr 
CMa\_249.0-14.0        &   248.986        &   -13.997        &     0.146     $\pm$    0.011        &     -0.07        &  7194        &   377     $\pm$   49        &     8     $\pm$    3
           \cr 
CMa\_249.0- 8.0        &   248.996        &    -7.999        &     0.286     $\pm$    0.016        &     -0.07        & 15215        &  1642     $\pm$   99        &    77     $\pm$    8
           \cr 
CMa\_251.9- 2.8        &   251.904        &    -2.837        &     1.063     $\pm$    0.066        &    \nodata      &  3493        & \nodata          &    \nodata
           \cr 
CMa\_252.0- 8.0        &   251.965        &    -8.004        &     0.296     $\pm$    0.025        &     -0.05        & 15333        &  1608     $\pm$   95        &    54     $\pm$    7
           \cr 
CMa\_252.0-15.0        &   251.977        &   -15.004        &     0.119     $\pm$    0.004        &     \nodata       &  5118        &    -9     $\pm$   67        &     2     $\pm$    1
           \cr 
\enddata
\tablenotetext{a}{(1) Field ID; (2) and (3) Galactic longitude and
  latitude respectively; (4) Median differential extinction
 and standard deviation: $^{\rm a}$ For this, data was taken from the
 Schlegel et al. (1998) dust maps,  modified using the correction from
 Bonifacio et al. (2000; their Eqn.\,1). 
 (5) The additional colour de-reddening, detailed in
 Sec.~\ref{extinc};  (6) 
 Number of stars in the CMD; (7)
 Number of CMa main sequence stars determined using the method described in
 Sec.~\ref{density1a}; (8) Number of young MS stars  and the rms uncertainty,
 estimated inside in the young MS extraction
 box outlined in Fig.~\ref{FigStarCnts_MS_ref_box_in_cmd}.
  There is no N$_{\rm Old\,\,MS}$ or N$_{\rm Young\,\,MS}$ 
 estimate if E(B-V)$_{\rm SFD98*}$ 
 $>$ 0.3\,mag. We note that fields with negative
  N$_{\rm Old\,\,MS}$ estimates are set to zero for the density profiles in
 Fig.~\ref{Fig_MS_BP_RC_profiles} and the surface density map in
 Fig.~\ref{Figcountprofile}. 
}\label{survey_params}
 \end{deluxetable}

\clearpage

\begin{deluxetable}{lccclcll}
\tabletypesize{\scriptsize}
\tablecaption{Target information. Continued}
\tablewidth{140mm}
\tablehead{  
\colhead{ID} & \colhead{l} & \colhead{b}   &
\colhead{E(B-V)$_{\rm SFD*}^{\rm a}$ $\pm$ $\sigma$}  & 
\colhead{$\Delta$ (B-R)$_0$}  & 
\colhead{N$_{\rm CMD}$ } & 
\colhead{N$_{\rm Old\,\,MS}$ $\pm$  $\sigma$} & 
\colhead{N$_{\rm Young\,\,MS}$ $\pm$$\sigma$} \cr
\colhead{} & \colhead{} & \colhead{} & \colhead{} &
  \colhead{}  & \colhead{}  & \colhead{} & \colhead{} \cr
  \colhead{} & \colhead{(deg)} & \colhead{(deg)} &
  \colhead{(mag)}    & \colhead{(mag)}    & \colhead{}  & \colhead{} & 
  \colhead{}   \cr 
\colhead{(1)} & \colhead{(2)} & \colhead{(3)} &
  \colhead{(4)}  & \colhead{(5)} & \colhead{(6)}  & \colhead{(7)} & \colhead{(8)}  \cr 
}
\startdata
CMa\_255.0- 3.0        &   254.974        &    -2.969        &     0.959     $\pm$    0.064        &    \nodata       &  8660        & \nodata       &    \nodata
           \cr 
CMa\_255.0-15.0        &   254.986        &   -15.006        &     0.118     $\pm$    0.005        &     \nodata       &  5164        &    -7     $\pm$   65        &     1     $\pm$    1
           \cr 
CMa\_255.0- 8.0        &   254.988        &    -8.001        &     0.307     $\pm$    0.041        &    \nodata        & 15506        & \nodata        &  \nodata
           \cr 
CMa\_257.9- 3.1        &   257.886        &    -3.099        &     1.566     $\pm$    0.056        &      \nodata        &  2212        & \nodata         &     \nodata
           \cr 
CMa\_258.0-15.0        &   257.979        &   -15.001        &     0.152     $\pm$    0.011        &    \nodata      &  5884        &  -141     $\pm$  104        &     6     $\pm$    2
           \cr 
CMa\_258.0- 8.0        &   257.985        &    -7.979        &     0.260     $\pm$    0.029        &     -0.08        & 18441        &  1677     $\pm$  146        &    45     $\pm$    6
\enddata
\tablenotetext{a}{(1) Field ID; (2) and (3) Galactic longitude and
  latitude respectively; (4) Median differential extinction
 and standard deviation: $^{\rm a}$ For this, data was taken from the
 Schlegel et al. (1998) dust maps,  modified using the correction from
 Bonifacio et al. (2000; their Eqn.\,1). 
 (5) The additional colour de-reddening, detailed in
 Sec.~\ref{extinc};  (6) 
 Number of stars in the CMD; (7)
 Number of CMa main sequence stars determined using the method described in
 Sec.~\ref{density1a}; (8) Number of young MS stars  and the rms uncertainty,
 estimated inside in the young MS extraction
 box outlined in Fig.~\ref{FigStarCnts_MS_ref_box_in_cmd}.
  There is no N$_{\rm Old\,\,MS}$ or N$_{\rm Young\,\,MS}$ 
 estimate if E(B-V)$_{\rm SFD98*}$ 
 $>$ 0.3\,mag. We note that fields with negative
  N$_{\rm Old\,\,MS}$ estimates are set to zero for the density profiles in
 Fig.~\ref{Fig_MS_BP_RC_profiles} and the surface density map in
 Fig.~\ref{Figcountprofile}. 
}\label{survey_params2}
 \end{deluxetable}

\clearpage





\end{document}